\documentclass[10pt,leqno]{amsart}

\usepackage{graphicx}
\usepackage{indentfirst,csquotes}

\usepackage[linesnumbered,ruled,vlined]{algorithm2e}

\usepackage{amssymb,amsthm,amsmath}
\usepackage{subcaption}
\usepackage{xcolor,paralist,hyperref,titlesec,fancyhdr,etoolbox}
\usepackage{multirow}
\newtheorem{theorem}{Theorem}[]

\newtheorem{proposition}[theorem]{Proposition}

\titleformat{\section}{\normalfont\Large\bfseries}{\thesection}{1em}{}
\titleformat{\subsection}{\normalfont\large\bfseries}{\thesubsection}{1em}{}

\titlespacing*{\section}{0pt}{0ex}{0ex}

\hypersetup{ colorlinks=true, linkcolor=black, filecolor=black, urlcolor=black }

\usepackage{lipsum}

\begin{document}
\title{Interpretable Small Training Set Image Segmentation Network Originated from Multi-Grid Variational Model} 
\author{Junying Meng}\thanks{Laboratory of Mathematics and Complex Systems (Ministry of Education of China),
School of Mathematical Sciences, Beijing Normal University, Beijing, 100875, China (junyingmeng@mail.bnu.edu.cn)}
\author{Weihong Guo}\thanks{Department of Mathematics, Applied Mathematics and Statistics, Case Western Reserve University, Cleveland, OH, 44106, USA (wxg49@case.edu)}
\author{Jun Liu}\thanks{Laboratory of Mathematics and Complex Systems (Ministry of Education of China),
School of Mathematical Sciences, Beijing Normal University, Beijing, 100875, China (jliu@bnu.edu.cn)}
\author{Mingrui Yang}\thanks{Lerner Research Institute, Cleveland Clinic, Cleveland, OH, 44195, USA (yangm@ccf.org)}

\maketitle

\begin{abstract}
The main objective of image segmentation is to divide an image into homogeneous regions for further analysis. This is a significant and crucial task in many applications such as medical imaging. Deep learning (DL) methods have been proposed and widely used for image segmentation. However, these  methods usually require a large amount of manually segmented data as training data and suffer from poor interpretability (known as the black box problem).
The classical Mumford-Shah (MS) model is effective for segmentation and provides a piece-wise smooth approximation of the original image. In this paper, we replace the  hand-crafted regularity term in the MS model with a data adaptive generalized learnable regularity term and use a multi-grid framework to unroll the MS model and obtain a variational model-based segmentation network  with better generalizability and interpretability. This approach allows for the incorporation of learnable prior information into the network structure design. Moreover, the multi-grid framework enables multi-scale feature extraction and offers a mathematical explanation for the effectiveness of the U-shaped network structure in producing good image segmentation results. Due to the proposed network originates from a variational model, it can also handle small training sizes. Our experiments on the REFUGE dataset, the White Blood Cell image dataset, and 3D thigh muscle magnetic resonance (MR) images demonstrate that even with smaller training datasets, our method yields better segmentation results compared to related state of the art  segmentation methods.
\end{abstract} 
\bigskip
\section{Introduction}
Image segmentation is a fundamental and critical task in the field of computer vision and has various important applications, particularly in the medical imaging domain. Medical image segmentation involves extracting regions of interest from image data, which plays an important role in assisting disease diagnosis and treatment efficacy assessment. For example, we can evaluate musculoskeletal diseases such as sarcopenia and osteoarthritis from thigh muscle composition. Thigh muscle segmentation is an important step in quantifying muscle composition. 
The objective of image segmentation is to find a partition of an image into disjoint regions. 
Over the years, a large number of methods have been proposed to achieve this objective. Among them, model-based and learning-based methods are the two primary tools. 

The model-based method is widely used and has achieved great success. Model-based approaches have natural mathematical interpretability as they directly translate expectation of results into models.  The variational method is one of the most commonly used model-based methods and has been proven to be accurate by incorporating prior information.
Representative works in the variational framework include Mumford-Shan \cite{mumford1989optimal}, Chan-Vese \cite{chan2001active}, geodesic active contours models \cite{caselles1997geodesic}, and so on. Segmentation models in the variational framework usually contain two terms. One of them is the fidelity term, which incorporates features of image pixels, such as gray level and spatial texture. Another term called the regularization term is contained in the segmentation model to enforce the spatial priors, such as spatial regularization, to ensure that the solutions of the model belong to a proper function space.  This regularization term helps to transform the original ill-posed problem into a well-posed problem, enabling accurate segmentation.

Total variation (TV) regularization is a well-known method widely used in image segmentation models due to its high accuracy. However, it is difficult to solve due to its non-smoothness, despite the development of many fast algorithms \cite{wang2008new} \cite{goldstein2009split} \cite{huang2008fast}.
An alternative method is the Threshold Dynamics (TD) method \cite{merriman1992diffusion} \cite{merriman1994motion},  which has similar regularization effects as TV but with higher computational efficiency.
TD was developed by Merriman, Bence, and Osher \cite{merriman1992diffusion} \cite{merriman1994motion} (also called MBO scheme) to approximate the motion of an interface driven by its mean curvature flow.  It can converge to motion by mean curvature (MMC) \cite{evans1993convergence}. 
TD has gained much attention for its stability and simplicity and has been extended to various applications, such as multi-phase flow with arbitrary surface tension \cite{2015threshold}, and multi-phase image segmentation \cite{tai2007image}. TD method is also extended to other applications such as image processing \cite{esedog2006threshold} \cite{wang2017efficient} \cite{merkurjev2013mbo} \cite{liu2022deep}, graph cut and data clustering \cite{van2014mean}, anisotropic interface motion problem \cite{merriman2000convolution}, etc.

For some challenging segmentation tasks that involve low contrast or missing edges, model-based methods may have limitations. For instance, medical image segmentation can be a challenging and time-consuming task due to various unavoidable interference factors, such as the presence of noise, intensity inhomogeneity, and similar textures and intensities of different objects of interest. 
In particular, thigh muscle segmentation is a complex task as different muscles are tightly packed together, and there are often no clear boundaries between them. 
To address such challenges, some methods have been proposed that incorporate auxiliary information as priors for medical segmentation. These semi-automatic methods, such as \cite{ahmad2014atlas} \cite{jolivet2014skeletal} \cite{molaie2020knowledge}, \cite{ogier2020novel}, \cite{guomarker} and \cite{ogier2017individual} have shown some efficiency in solving these problems. However, these methods require manual intervention, which can be time-consuming and tedious. In the past few years, several automatic methods have been proposed \cite{andrews2015generalized} \cite{baudin2012prior} \cite{chen2003using} \cite{kemnitz2017validation} \cite{lotjonen2010fast} \cite{le2016volume} \cite{mesbah2019novel} \cite{sharma2019mammogram} \cite{yokota2018automated}. Since intensity-based methods cannot distinguish different regions, most automatic methods are based on shape-based methods.
Joint image processing models have attracted much attention in recent years. A segmentation model with adaptive priors from joint registration has been proposed in \cite{li2022image}. In this approach, registration plays the role of providing a shape prior to guiding the segmentation process. This method has been shown to outperform separate segmentation and registration methods as well as other joint methods. However, this method is highly dependent on the selection of moving images used for registration. When there is a significant difference between the moving image and the target image, the shape prior provided by the registration may not be accurate enough, leading to poor segmentation results. Moreover, this method requires operating on independent image pairs, and the computational cost of the energy function minimization is high.

Though model-based methods are mathematically interpretable, the accuracy of the results depends  heavily on the choice of hand-crafted regularization term, which typically requires extensive parameter tuning and can be a laborious and time-consuming process.
 
In recent years, data-driven learning-based methods have gained popularity in image processing tasks and have shown remarkable performance improvements due to their ability to extract good features automatically. 
These methods have been successfully applied in various image processing tasks such as image segmentation, denoising, super-resolution, and so on. Unlike model-based methods that rely on handcrafted priors, learning-based methods derive implicit priors from a large amount of data by optimizing a loss function on training image pairs. The trained network can be used on test images without requiring any parameter tuning. Lots of 2D Convolutional Neural Networks (CNNs) have been designed for image segmentation.
 Full convolution network (FCN) uses a multiscale and deep feature fusion approach to achieve dense pixel prediction and has been adopted in many advanced segmentation approaches,  such as \cite{liu2015semantic} \cite{ghiasi2016laplacian} \cite{lin2016efficient} \cite{lin2017exploring} \cite{sanderson2022fcn}. 
Another important semantic segmentation network is the U-Net \cite{ronneberger2015u}, which is mainly composed of an Encoder-Decoder structure with skip connections to better reconstruct intrinsic features at multiple scales. 
DeepLabV3+ \cite{chen2018encoder} is another state-of-the-art deep convolutional neural network architecture for semantic image segmentation, which combines the advantages of the atrous spatial pyramid pooling (ASPP) module and the encoder-decoder structure. Due to its great performance in many benchmark datasets, DeepLabV3+ has become the baseline for the semantic segmentation DCNNs.
R2U-Net \cite{alom2018recurrent} is an extension of the U-Net architecture that utilizes recurrent layers and residual connections to improve the accuracy of segmentation, which has promising results in various medical image segmentation tasks.

The above 2D image segmentation network takes 2D data as input and produces the 2D output, which only learns 2D spatial features. For some medical images such as X-rays, 2D image segmentation networks can effectively perform segmentation due to the information being constrained in a 2D plane. For some 3D volumetric data such as CT and MRI scans,  the inter-slice information is important when the region of interest is distributed across multiple slices. However, 2D convolutions may not be able to learn the inter-slice correction information, which can affect the segmentation performance \cite{niyas2022medical}. So some 3D segmentation networks use 3D convolution kernels proposed to learn the inter-slice information and provide better segmentation performance.
By replacing all 2D operations in U-Net \cite{ronneberger2015u} with 3D operations, a 3D CNN network 3D U-Net \cite{cciccek20163d} is proposed for volume segmentation, which shows significant improvement in segmentation results compared to 2D CNN.
Chen et al. proposed Med3D \cite{chen2019med3d}, a 3D network that develops multiple pre-trained 3D medical image segmentation models by jointly training on multiple domain 3D datasets.  Med3D is considered to be a suitable choice for small 3D image datasets. With a 3D fully convolutional architecture,  DenseVoxNet \cite{yu2017automatic} is capable of achieving effective volume-to-volume prediction. Compared to other 3D CNNs, this network can achieve better segmentation performance with fewer parameters.

Many learning-based image segmentation neural networks are designed without fully considering the superiority of model-based methods. Furthermore, all of these deep learning models are known for their "black box" nature, which lacks model interpretability. Additionally, these method rely on large training datasets that are difficult to obtain. Training such models on small datasets can lead to overfitting issues, resulting in biased models with poor generalizability to unseen clinical datasets. Recently, researchers have attempted to build semi-supervised \cite{you2022simcvd, luo2022semi} or unsupervised \cite{papandreou2015weakly} deep learning models to address the issue of limited training data. However, these models typically perform worse than conventional deep learning models trained with sufficient data. Therefore, there is an unmet need to develop fast, accurate, and reliable segmentation models that can be trained with limited data and have good interpretability to build reliable clinical tools for clinical applications.

Unrolling methods combine the benefits of both the model-based methods and the learning-based methods. These methods involve designing networks by unfolding iterative algorithms used to solve the models, providing a mathematical interpretation based on the model-based method. More researchers are paying more attention to algorithm unrolling. For example, one approach involves unrolling the MF iterations of CRF into an RNN, which is then concatenated with the semantic segmentation network to form a network \cite{zheng2015conditional}. 
The concatenated network is similar to traditional semantic segmentation followed by CRF-based post-processing, while end-to-end training can be performed across the entire network. The deep parsing network, developed by \cite{liu2017deep}, follows a similar approach.  
MgNet \cite{he2019mgnet} developed an Encoder-Decoder framework for CNN architectures that relies on the multi-grid algorithm for solving PDEs. It is possible to construct some networks with U-shaped architecture using multi-grid numerical schemes. These unrolling methods allow for the combination of model-based and learning-based methods, resulting in networks that are both interpretable and effective.

We start with the classical Mumford-Shah model and replace the conventional hand-crafted regularity term in MS model with a more flexible and generalized learnable regularity term and approximate the arc length regularity by a threshold dynamics.
By unrolling such a variational model, we get a learnable segmentation network that is based on a solid and general mathematical foundation.  
Unlike most unrolling methods that rely on a single scale for feature extraction,  we employ a multi-grid framework for unrolling, which leads to the similar multi-scale feature learning process to the current state-of-the-art neural network, U-Net. The proposed method combines the advantages of model-based and learning-based methods, resulting in higher interpretability and generality.

The main contributions of this work are the following:
\begin{itemize}
\item{
By unrolling the iterative algorithm of solving the modified Mumford-Shan model with a generalized learnable regularity term, we get a learnable segmentation network that has mathematical interpretability and generalizability. }
\end{itemize}

\begin{itemize}
\item{An encoder-decoder network architecture is designed using the multi-grid framework to achieve multi-scale features extraction. This approach provides a mathematical explanation for the effectiveness of U-shaped networks.
}
\end{itemize}

\begin{itemize}
\item{Since the proposed network originates from a variational model, it can handle small training sizes, which is ideal for medical image segmentation tasks that may face challenges in acquiring large training sets. Moreover,
by replacing the 2D operations in the proposed network with 3D operations, the proposed network can be expanded to a 3D network for 3D medical image segmentation. 
}
\end{itemize}

The rest of this paper is organized as follows. In Section \ref{relatedwork}, we review the most related works. The variational model is given in Section \ref{proposed}. The algorithm for the variational model is analyzed in Section \ref{algorithm} and  Section \ref{net} shows how to construct a segmentation network architecture according to the proposed variational method. The experimental results are shown in Section \ref{experiment}. Finally, we conclude this paper in Section \ref{conclusion}.

\section{Related works}\label{relatedwork}
\subsection{Mumford-Shah model}
The Mumford-Shah model \cite{mumford1989optimal} is a classical variational image segmentation model that was proposed by Mumford and Shah in 1989. Assuming that the image domain $\Omega$ is decomposed into $N$ disjoint connected open subsets $\Omega_{n}$ with each subset $\Omega_{n}$ having a piece-wise smooth boundary and $\Gamma$ be the union of the boundaries of the $\Omega_{n}$ inside $\Omega$, so that
\begin{equation}\label{equ2.1}
\Omega = \Omega_{1}\cup ... \cup \Omega_{N} \cup \Gamma,
\end{equation}
the main idea of the Mumford-Shah model is to find the boundary $\Gamma$ and $u$, such that $u$ is a nearly piece-wise smooth approximation of $f$ and $\Gamma$ is a set of edges. The Mumford-Shah model has the following formulation:
\begin{equation*}
\mathop{min}_{u\in H^1, \Gamma}\Big\{\frac{1}{2}\int_{\Omega}(f(x)-u(x))^{2}dx +\frac{\lambda}{2}\int_{\Omega\backslash \Gamma}|\nabla u(x)|^{2}dx +\alpha|\Gamma|\Big\},
\end{equation*}
where $f :\Omega \rightarrow \mathbb{R}$ is the input image, $\lambda, \alpha>0$ are parameters, and $|\Gamma|=\sum_{n=1}^{N}|\partial \Omega_{n}|$ represents the length of the close subsets $\Gamma \subset \Omega$.  The first term is the fidelity term, which ensures that $u$ is not much deviation from the real image $f$. The second term is a regularization term such that $u\in H^{1}(\Omega\backslash \Gamma)$ is a nearly piece-wise smooth approximation of $f$
By solving this model, a piece-wise smooth approximation of the original image can be obtained with the shortest boundary length. The Mumford-Shah model remains an important foundation for the development of image segmentation techniques and continues to drive new research in this field.

\subsection{Multi-grid algorithm}
The multi-grid algorithm \cite{henson2003multigrid} is a computational method utilized to solve numerical problems, particularly those associated with partial differential equations. The efficiency of multi-grid algorithm has been demonstrated in  \cite{terzopoulos1986image} for low-level computer vision problems and in \cite{chen2007nonlinear}  for variational image processing tasks. 
Many linear problems of the form $A\mu = b$ can be solved using multigrid methods, which have become a widely used approach. Let $\nu$ be an approximation of the exact solution $\mu$ and e be the error, i.e., $e = \mu - \nu$. 
By defining the residual as $r = b - A\nu,$ then the residual equation $Ae = r$ can be obtained.

The multi-grid method consists of two basic parts: relaxation iteration and coarse grid correction. Some common relaxation operations including Gauss-Seidel iteration and relaxation iteration are used to reduce the error and make the solution converge. 
Coarse grid correction refers to compressing the current grid into a coarser grid, then the residual $r^{h}$ on a fine grid is limited to a coarser grid $r^{h+1} = I_{h}^{h+1}r^{h}$. In this process, the restriction operator $I_h^{h+1}$ is used to map the fine grid to the coarse grid. It achieves this by averaging or aggregating the values from the fine grid to obtain corresponding values on the coarse grid. The limited residual is then utilized as the right-hand side of the residual equation on the coarser grid, i.e.
$$A^{h+1}e^{h+1}=r^{h+1},$$
where $A^{h+1}$ represents the linear operator on the coarser grid.
The solution of the above function is then interpolated to the fine grid to correct the fine-grid approximation,
$$\nu^{h}\leftarrow \nu^{h}+I_{h+1}^{h}e^{h+1}.$$
Here, The prolongation operator $I^{h}_{h+1}$ maps the coarse grid back to the fine grid by interpolating or extending values from the coarse grid to estimate corresponding values on the fine grid.
Then a multi-grid algorithm is defined by solving the coarse-grid equation recursively using this two-grid process.

For the nonlinear algebraic equations $A(\mu)=b$. Similarly, the error can be defined as $e = \mu - \nu$, and now the residual $r = b - A(\nu)$. Then
\begin{equation}\label{equ2.2}
A(\mu)-A(\nu)=r.
\end{equation}
Due to the nonlinearity of $A$, $A(e)\neq r$, which means that for nonlinear equations, we cannot obtain the error by solving a linear equation on the coarse grid as in the standard multi-grid method mentioned above. Therefore, equation (\ref{equ2.2}) is utilized as the residual equation instead. The Full Approximation Scheme (FAS), is presented in \cite{henson2003multigrid}, where it directly applies the multi-grid method to the original equation $A(\mu) = b$ and employs the nonlinear residual equation (\ref{equ2.2}) as the basis for coarse-grid correction.
Specifically, assume $\nu^{h}$ is an approximation for
\begin{equation}\label{equ2.3}
 A^{h}(\mu^{h})=b^{h}
 \end{equation}
 on the fine-grid, then $e^{h}=\mu^{h}-\nu^{h}$.
The residual equation (\ref{equ2.2}) on the coarse grid can be expressed as:
$$A^{h+1}(\nu^{h+1}+e^{h+1})-A^{h+1}(\nu^{h+1}) = r^{h+1}.$$
According to $r^{h+1} = I_{h}^{h+1}r^{h} = I_{h}^{h+1}(b^{h}-A^{h}(\nu^{h}))$ and $\nu^{h+1} = I_{h}^{h+1}\nu^{h}$, the coarse-grid residual equation can be written as
$$A^{h+1}(I_{h}^{h+1}\nu^{h}+e^{h+1})=A^{h+1}(I_{h}^{h+1}\nu^{h})+I_{h}^{h+1}(b^{h}-A^{h}(\nu^{h})).$$
Let $\mu^{h+1}:=I_{h}^{h+1}\nu^{h}+e^{h+1}$ and $b^{h+1}:= A^{h+1}(I_{h}^{h+1}\nu^{h})+I_{h}^{h+1}(b^{h}-A^{h}(\nu^{h}))$, we can obtain $A^{h+1}\mu^{h+1} = b^{h+1},$  which has the same form as the initial fine-grid equation (\ref{equ2.3}).
Assume $\mu^{h+1}$ is the solution of this equation, then the coarse-grid error $e^{h+1}= \mu^{h+1}-I_{h}^{h+1}\nu^{h}$ can be interpolated up to the fine grid and used to correct the fine-grid approximation $\nu^{h}$:
$$\nu^{h}\leftarrow \nu^{h}+I_{h+1}^{h}e^{h+1}.$$
The full approximation scheme (FAS) is so named because it solves the coarse-grid problem using the full approximation instead of the error $e^{h+1}$.

\section{The proposed model}\label{proposed}
Several studies have shown that piece-wise smooth models can effectively handle images with intensity inhomogeneity, as evidenced by findings from \cite{tsai2001curve} and \cite{vese2002multiphase}. We construct an image segmentation network based on an improvement to the Mumford-Shan model. We propose a novel approach to improve the classical Mumford-Shah (MS) model for image segmentation. The MS model is effective for segmentation tasks and provides a piece-wise smooth approximation of the original image. However, the regularity terms in the MS model are pre-determined and thus limit its generalizability and adaptivity to different image datasets.
To address this issue, we propose replacing the hand-crafted regularity terms in the MS model with  data adaptive generalized learnable regularity terms. This means that the regularity terms are learned from the data and can adapt to different image datasets. By doing so, we obtain a more adaptive and flexible model that can better capture the characteristics of the image data and improve the segmentation performance.

Let $f: \Omega \rightarrow  \mathbb{R}$ be the image function. In order to better extract and describe image features, we can map the original image space $\mathbb{F}$ to a high-dimensional feature space $\hat{\mathbb{F}}$ through a head network to be learned. This head network can be represented as a feature extractor operator $\mathcal{G}: \mathbb{F}\rightarrow\hat{\mathbb{F}}$. Therefore, we can define the new feature representation of $f$ as $\hat{f}=\mathcal{G}(f)$, where $\hat{f} = \{\hat{f}_{1},...,\hat{f}_{I}\}:\Omega\rightarrow\mathbb{R}^{I}$, and $\hat{f}_{i}$ represents the i-th feature for $i\in\{1,...,I\}$.

The Mumford-Shah model for segmenting the feature $\hat{f}_{i}$ can be expressed as:
\begin{equation}\label{equ3.1}
\mathop{min}_{\hat{u}_{i}, \Gamma_{i}}\Big\{\frac{1}{2}\int_{\Omega}(\hat{f}_{i}(x)-\hat{u}_{i}(x))^{2}dx +\frac{\lambda}{2}\int_{\Omega\backslash \Gamma}|\nabla \hat{u}_{i}(x)|^{2}dx +\alpha|\Gamma_{i}|\Big\},
\end{equation}
where the boundary $\Gamma_{i}$ separates $\Omega$ into $N$ disjoint open subsets $\Omega_{i,n}$, i.e. $\Omega=\Omega_{i,1}\cup \Omega_{i,2}...\cup\Omega_{i,N}\cup\Gamma_{i}$. $\hat{f}_{i}$ is approximated by a smooth function $\hat{u}_{i,n}$ in each $\Omega_{i,n}$, then the above model can be reformulated as follows:
$$\mathop{min}_{\{\hat{u}_{i,n}, \Omega_{i,n}\}_{n=1}^{N}}\Big\{\sum_{n=1}^{N}\big(\frac{1}{2}\int_{\Omega_{i,n}}(\hat{f}_{i}(x)-\hat{u}_{i,n}(x))^{2}dx+\frac{\lambda}{2}\int_{\Omega_{i,n}}|\nabla\hat{u}_{i,n}(x)|^{2}dx +\alpha|\partial \Omega_{i,n}|\big)\Big\}.$$
To compute the optimal partitions, let $\hat{v}_{i,n}(x), n=1,...,N,$ denote the characteristic functions of the disjoint subdomains $\Omega_{i,n}$, i.e.
\begin{equation*}
\hat{v}_{i,n}(x)=I_{\Omega_{i,n}}(x): =\left\{
\begin{aligned}
1, x\in\Omega_{i,n},\\
0, x\notin \Omega_{i,n},
\end{aligned}
\right. n = 1,...,N.
\end{equation*}
Then the boundary length of the disjoint subdomains can be expressed as:
$$|\partial\Omega_{i,n}|=\int_{\Omega}|\nabla \hat{v}_{i,n}(x)|dx, n =1,...,N.$$
Now the Mumford-Shah model can be written as
\begin{equation*}
\begin{aligned}
\mathop{min}_{\{\hat{u}_{i,n}, \hat{v}_{i,n}\}_{n=1}^{N}}\Big\{&\sum_{n=1}^{N}\big(\frac{1}{2}\int_{\Omega}(\hat{f}_{i}(x)-\hat{u}_{i,n}(x))^{2}\hat{v}_{i,n}(x)dx+\frac{\lambda}{2}\int_{\Omega}|\nabla \hat{u}_{i,n}(x)|^{2}\hat{v}_{i,n}(x)dx\\&+\alpha\int_{\Omega}|\nabla\hat{v}_{i,n}(x)|dx\big)\Big\},
\end{aligned}
\end{equation*}
subject to  $$\sum_{n=1}^{N}\hat{v}_{i,n}(x)=1, \forall x\in\Omega.$$
The above problem is not convex so instead of restricting $\hat{v}_{i,n}(x)$ to be 0 or 1, we relax $\hat{v}_{i,n}(x)$ to be between 0 and 1, i.e., $\hat{v}_{i,n}(x)$ is a relaxed indicator function of $\Omega_{i,n}$.   Then the segmentation condition can be written as the following simplex:
\begin{equation}\label{equ3.2}
\mathbb{V}=\{\hat{v}_{i}=(\hat{v}_{i,1}, ... , \hat{v}_{i,N})\in[0,1]^{N}:\sum_{n=1}^{N}\hat{v}_{i,n}(x)=1, \forall x\in\Omega\}.
\end{equation}
By incorporating the segmentation condition (\ref{equ3.2}), the relaxed Mumford-Shah model is as follows:
\begin{equation*}
\begin{aligned}
\mathop{min}_{\{\hat{u}_{i,n}\}_{n=1}^{N},\hat{v}_{i}\in\mathbb{V}}\Big\{&\sum_{n=1}^{N}\big(\frac{1}{2}\int_{\Omega}(\hat{f}_{i}(x)-\hat{u}_{i,n}(x))^{2}\hat{v}_{i,n}(x)dx+\frac{\lambda}{2}\int_{\Omega}|\nabla \hat{u}_{i,n}(x)|^{2}\hat{v}_{i,n}(x)dx\\ &+\alpha\int_{\Omega}|\nabla \hat{v}_{i,n}(x)|dx\big)\Big\}.
\end{aligned}
\end{equation*}
The final term in the model is a TV regularization, which represents the length $|\partial\Omega_{i,n}|$. 
Since TV is non-smooth, a smooth regularization term can be used to approximate the length $|\partial\Omega_{i,n}|$. This can be achieved using the threshold dynamics method \cite{2015threshold}, \cite{merriman1994motion}:
$$|\partial\Omega_{i,n}| \approx \sqrt{\frac{\pi}{\sigma}}\sum_{\hat{n}=1,\hat{n}\neq n}^{N}\int_{\Omega}\hat{v}_{i,n}(x)(k*\hat{v}_{i,\hat{n}}(x))dx,$$
where the symbol $*$ represents the convolution operator, and $k$ is a Gaussian kernel with $k(x)=\frac{1}{2\pi\sigma^{2}}e^{-\frac{|x|^{2}} {2\sigma^{2}}}$. This regularization term penalizes isolated pixels and imposes spatial dependencies on the model. It has been shown that when $\sigma\rightarrow 0$, this threshold dynamics regularization converges to $|\partial\Omega_{i,n}|$ in the sense of $\Gamma$-convergence \cite{miranda2007short}.

The entropy term enables the segmentation to be a soft threshold. This is beneficial to backpropagation in the network since it can ensure the solution to be smooth. Therefore, we provide the following modified Mumford-Shah model, incorporating the introduced entropy term:
\begin{equation*}
\begin{aligned}
\mathop{min}_{\{\hat{u}_{i,n}\}_{n=1}^{N}, \hat{v}_{i}\in\mathbb{V}}&\Big\{\sum_{n=1}^{N}\Big(\frac{1}{2}\int_{\Omega}(\hat{f}_{i}(x)-\hat{u}_{i,n}(x))^{2}\hat{v}_{i,n}(x)dx + \frac{\lambda}{2}\int_{\Omega}\hat{v}_{i,n}(x)|\nabla \hat{u}_{i,n}(x)|^{2}dx  \\
&+\alpha \int_{\Omega}e(x)\hat{v}_{i,n}(x)\big(k*(1-\hat{v}_{i,n})\big)(x)dx+\varepsilon\int_{\Omega}\hat{v}_{i,n}(x)\ln \hat{v}_{i,n}(x)dx\Big)\Big\}.
\end{aligned}
\end{equation*}
For the third term, define $\mathcal{R}(\hat{v}_{i}):= \alpha\langle e\hat{v}_{i}, k*(1-v_{i})\rangle= \alpha\sum_{n=1}^{N}\int_{\Omega}e(x)\hat{v}_{i,n}(x)(k*(1-\hat{v}_{i,n}))(x)dx.$
Here $e \geq 0$ is a weighting function such as edge detection function $e(x)=\frac{1}{1+\gamma\|\bigtriangledown \hat{f}_{i}(x)\|}$, where $\gamma>0$. When the kernel $k$ satisfies certain conditions, it has been demonstrated in \cite{liu2011fast} that $\mathcal{R}(\hat{v}_{i})\propto \sum_{n=1}^{N}\int_{\partial \Omega_{i,n}}e \hspace{0.1cm} ds$.  In particular, when $e(x)\equiv 1$, the penalty term $\mathcal{R}(\hat{v}_{i})$ serves to penalize the length of boundaries.

When the kernel function $k$ is chosen as a semi-positive definite, $\mathcal{R}$ is concave. To this end, we can replace the concave functional $\mathcal{R}$ with its supporting hyperplane, given by
$$\mathcal{R}(\hat{v}_{i})=\mathcal{R}(\hat{v}_{i}^{t_{1}})+\langle p_{i}^{t_{1}}, \hat{v}_{i}-\hat{v}_{i}^{t_{1}}\rangle,$$
where $p_{i}^{t_{1}}=\alpha ((k*(1-\hat{v}_{i}^{t_{1}}))e-k*(e \hat{v}_{i}^{t_{1}}))\in\partial \mathcal{R}(\hat{v}_{i}^{t_{1}})$ and $\partial \mathcal{R}(\hat{v}_{i}^{t_{1}})$ is the subgradient of the concave functional $\mathcal{R}$ at $\hat{v}_{i}^{t_{1}}$. In the special case where $e=1$, we have $p_{i}^{t_{1}} = \alpha k*(1-2\hat{v}_{i}^{t_{1}}).$
Now the modified Mumford-Shah model can be written as 
\begin{equation}\label{equ3.3}
\mathop{min}_{\{\hat{u}_{i,n}\}_{n=1}^{N}, \hat{v}_{i}\in\mathbb{V}}\mathcal{J}(\hat{u}_{i,n},\hat{v}_{i,n}), 
\end{equation}
where
\begin{equation}\label{equ3.4}
\begin{aligned}
\mathcal{J}(\hat{u}_{i,n},\hat{v}_{i,n}) =& \sum_{n=1}^{N}\Big(\frac{1}{2}\int_{\Omega}(\hat{f}_{i}(x)-\hat{u}_{i,n}(x))^{2}\hat{v}_{i,n}(x)dx + \frac{\lambda}{2}\int_{\Omega}\hat{v}_{i,n}(x)|\nabla \hat{u}_{i,n}(x)|^{2}dx \\
&+\alpha\int_{\Omega}k*(1-2\hat{v}_{i,n}^{t_{1}}(x))(\hat{v}_{i,n}(x)-\hat{v}_{i,n}^{t_{1}}(x))dx+\varepsilon\int_{\Omega}\hat{v}_{i,n}(x)\ln \hat{v}_{i,n}(x)dx\\
&+ \alpha\int_{\Omega}\hat{v}_{i,n}^{t_{1}}(x)(k*(1-\hat{v}_{i,n}^{t_{1}}))(x)dx\Big).
\end{aligned}
\end{equation}
The predetermined regularity term in the above model impose limitations on its ability to generalize and adapt to various image datasets. To overcome this challenge, we propose substituting the manually designed regularity term in the above model with a data-adaptive generalized learnable regularity term. This approach enhances the model's flexibility and  adaptability, allowing it to better handle diverse image datasets. Thus, by introducing the learnable operator $\mathcal{L}$, we get the general model: 
\begin{equation}\label{equ3.5}
\begin{aligned}
\mathcal{J}(\hat{u}_{i,n},\hat{v}_{i,n}) =& \sum_{n=1}^{N}\Big(\frac{1}{2}\int_{\Omega}(\hat{f}_{i}(x)-\hat{u}_{i,n}(x))^{2}\hat{v}_{i,n}(x)dx + \frac{\lambda}{2}\int_{\Omega}\hat{v}_{i,n}(x)|\mathcal{L}\hat{u}_{i,n}(x)|^{2}dx \\
&+\alpha\int_{\Omega}k*(1-2\hat{v}_{i,n}^{t_{1}}(x))(\hat{v}_{i,n}(x)-\hat{v}_{i,n}^{t_{1}}(x))dx+\varepsilon\int_{\Omega}\hat{v}_{i,n}(x)\ln \hat{v}_{i,n}(x)dx\\
&+ \alpha\int_{\Omega}\hat{v}_{i,n}^{t_{1}}(x)(k*(1-\hat{v}_{i,n}^{t_{1}}))(x)dx\Big).
\end{aligned}
\end{equation}
Compared to the original Mumford-Shah model, our modified model replaces the nonsmooth TV regularity term with a smooth regularity term based on threshold dynamics method and is more general with a data adaptive generalized learnable regularity term.

\section{The numerical scheme for constructing the network}\label{algorithm}
The model (\ref{equ3.5}) can be solved by an alternating direction minimization (ADMM) algorithm.

\begin{subequations}
\begin{align}
\hat{u}_{i,n}^{t_{1}+1} &= \mathop{\arg\min}_{\hat{u}_{i,n}}\mathcal{J}(\hat{u}_{i,n}, \hat{v}_{i,n}^{t_{1}}) \label{4.1} \\
v_{i,n}^{t_{1}+1} &=\mathop{\arg\min}_{\hat{v}_{i,n}}\mathcal{J}(\hat{u}_{i,n}^{t_{1}+1}, \hat{v}_{i,n}) \label{4.2}
\end{align}
\end{subequations}

\begin{proposition}\label{prop}
The sequence $(\hat{u}_{i,n}, \hat{v}_{i,n})$ generated by $(\ref{4.1})-(\ref{4.2})$ satisfies
$$\mathcal{J}(\hat{u}_{i,n}^{t_{1}+1},\hat{v}_{i,n}^{t_{1}+1})\leq \mathcal{J}(\hat{u}_{i,n}^{t_{1}},\hat{v}_{i,n}^{t_{1}}).$$
\end{proposition}
Proposition \ref{prop} is straightforward to prove and it shows that the cost decreases with each iteration of $(\ref{4.1})$ and $(\ref{4.2})$. Next, we will solve these subproblems.

For the $\hat{u}$-subproblem, we can consider the Euler-Lagrange equation for minimizing the following functional:
$$\sum_{n=1}^{N}\Big( \frac{1}{2} \int_{\Omega}(\hat{f}_{i}(x)-\hat{u}_{i,n}(x))^{2}\hat{v}_{i,n}^{t_{1}}(x)dx + \frac{\lambda}{2}\int_{\Omega}\hat{v}_{i,n}^{t_{1}}(x) |\mathcal{\mathcal{L}}\hat{u}_{i,n}(x)|^{2}dx \Big).$$
Then we have
\begin{equation}\label{equ4.3}
\Big(\hat{v}_{i,n}^{t_{1}}(x)+\lambda \mathcal{L}^{*}\hat{v}_{i,n}^{t_{1}}(x)\mathcal{L}\Big)\hat{u}_{i,n}(x)=\hat{v}_{i,n}^{t_{1}}(x)\hat{f}_{i}(x),
\end{equation}
where $\mathcal{L}^{*}$ is the conjugation of $\mathcal{L}$. It can be solved using a time-marching method by evolving function:
$$\frac{\partial{\hat{u}_{i,n}(x,t)}}{\partial t} = \hat{f}_{i}(x)\hat{v}_{i,n}^{t_{1}}(x)-\mathcal{A}(\hat{u}_{i,n}(x,t)),$$
where $\mathcal{A} =  \hat{v}_{i,n}^{t_{1}}(x) + \lambda\mathcal{L}^{*}\circ \hat{v}_{i,n}^{t_{1}}(x)\circ\mathcal{L}$. Then we get the iteration scheme:
\begin{equation}\label{equ4.4}
\hat{u}_{i,n}^{t_{1}+1}(x)=\hat{u}_{i,n}^{t_{1}}(x)+\triangle t\big(\hat{f}_{i}(x)\hat{v}_{i,n}^{t_{1}}(x)-\mathcal{A}\hat{u}_{i,n}^{t_{1}}(x)\big),
\end{equation}
where $\triangle t$ represents the step size.

For the $\hat{v}$-subproblem,
\begin{equation*}
\begin{aligned}
\hat{v}_{i,n}^{t_{1}+1}&=\mathop{\arg\min}_{v_{i,n}}\Big\{\sum_{n=1}^{N}\Big(\frac{1}{2}\int_{\Omega}(\hat{f}_{i}(x)-\hat{u}_{i,n}^{t_{1}+1}(x))^{2}\hat{v}_{i,n}(x)dx + \frac{\lambda}{2}\int_{\Omega}\hat{v}_{i,n}(x)|\mathcal{L}\hat{u}_{i,n}^{t_{1}+1}(x)|^{2}dx \\
&+\alpha \int_{\Omega}k*(1-2\hat{v}_{i,n}^{t_{1}})(x)(\hat{v}_{i,n}(x)-\hat{v}_{i,n}^{t_{1}}(x))dx+\varepsilon\int_{\Omega}\hat{v}_{i,n}(x)\ln \hat{v}_{i,n}(x)dx\\
&+ \alpha \int_{\Omega}\hat{v}_{i,n}^{t_{1}}(x)(k*(1-\hat{v}_{i,n}^{t_{1}}))(x)dx\Big)\Big\}.
\end{aligned}
\end{equation*}
This subproblem has a solution that can be obtained through the use of the first-order optimization condition, which yields a softmax function:
\begin{equation}\label{equ4.5}
\hat{v}_{i,n}^{t_{1}+1}(x)=\frac{e^{\frac{-\frac{1}{2}(\hat{f}_{i}(x)-\hat{u}_{i,n}^{t_{1}+1}(x))^{2}-\frac{\lambda}{2} |\mathcal{L}\hat{u}_{i,n}^{t_{1}+1}(x)|^{2}-\alpha k*(1-2\hat{v}_{i,n}^{t_{1}})(x)}{\varepsilon}}}
{\sum_{n=1}^{N}e^{\frac{-\frac{1}{2}(\hat{f}_{i}(x)-\hat{u}_{i,n}^{t_{1}+1}(x))^{2}-\frac{\lambda}{2}|\mathcal{L}\hat{u}_{i,n}^{t_{1}+1}(x)|^{2}-\alpha k*(1-2\hat{v}_{i,n}^{t_{1}})(x)}{\varepsilon}}}.
\end{equation}
The solution obtained from the above formula incorporates the spatial prior information represented by terms  $\alpha k*(1-2\hat{v}_{i,n})(x)$ and $\frac{\lambda}{2}|\mathcal{L}\hat{u}_{i,n}(x)|^{2}$, which will be integrated and learned within the designed network structure.

The process of segmenting images using model (\ref{equ3.5}) can be described by the following Algorithm \ref{algorithm4.1}:

\begin{algorithm}[H]
\SetAlgoNoLine
\caption{Image segmentation by estimating the deep feature}
\label{algorithm4.1}
\SetKwData{f}{f}
\SetKwData{u}{u}
\SetKwData{v}{v}
\SetKwData{T}{T}
\SetKwData{G}{G}
\textbf{Input:}  image $f$.\\
\textbf{Deep features:} $\hat{f}=\mathcal{G}(f),$ $\hat{u}_{i,n}^{0}=\hat{f}_{i}.$\\
\For{$ t_{1} = 0 : T-1$}{
    Features passing: updating $\hat{u}_{i,n}^{t_{1}+1}$ by (\ref{equ4.4})\;
    Attention parameter: updating $\hat{v}_{i,n}^{t_{1}+1}$ by (\ref{equ4.5})\;
}
\textbf{Output:}Segmentation function $\mathbf{v}=\bar{\mathcal{G}}(\hat{\mathbf{v}}^{T})$\;
\end{algorithm}

\subsection{Multi-grid solver}
The task of segmentation is formulated as an energy functional minimization problem, as shown in equation (\ref{equ3.5}). However, this energy functional is often computationally demanding to minimize, as it requires a large number of iterations to solve the $\hat{u}$-subproblem. Moreover, the feature extraction process described in Algorithm \ref{algorithm4.1} only considers information from a single scale, thereby lacking the incorporation of information from multiple scales.

As demonstrated in \cite{tai2002global}, the multi-grid method is a highly effective solver for both linear and nonlinear equation systems. This technique has been successfully applied in various low-level computer vision problems \cite{terzopoulos1986image}.
In image processing tasks, the multi-grid method has been utilized to solve variational models  \cite{chen2007nonlinear}. 
If the multi-grid method is used to solve optimization problems (\ref{equ4.3}), features can be extracted from different scales. These features can be processed using the similar process to the state-of-the-art neural network, U-Net. So it is possible to significantly improve the accuracy of the resulting segmentation results. In addition, it can improve the efficiency of image restoration. This is because multi-grid techniques can rapidly eliminate parameters in the model, facilitating faster information propagation.
In \cite{henson2003multigrid}, a Full Approximation Scheme (FAS) was introduced as an approach to apply the multi-grid method directly to nonlinear problems. Here (\ref{equ4.4}) can be seen as a relaxation scheme in FAS to ensure the smoothness of the error. 
By introducing the grid transfer operator, features can be transferred between the different scales.  A summary of the multi-grid solver for the proposed model is provided in Algorithm \ref{algorithm4.1}. The algorithm outlines the steps involved in solving problems at multiple scales, transferring features between coarse and fine grids, and refining the solution.

\begin{algorithm}[htbp]
\caption{$\hat{u}_{i,n},\hat{v}_{i,n}= $Multi-grid solver$(\hat{f}_{i}, H, T_{1}, T_{2}, ... , T_{H})$}\label{algorithm4.2}
\textbf{Input:}   $\hat{u}_{i,n}^{1,0}=\hat{f}_{i}^{1}=\hat{f}_{i}$, $\hat{v}_{i,n}^{1,0} = softmax(\hat{f}_{i})$, $\hat{F}_{i,n}^{1}=\hat{v}_{i,n}^{1,0}\hat{f}_{i}$, $A^{1} = \hat{v}_{i,n}^{1,0}I+\lambda\mathcal{L}^{*}\hat{v}_{i,n}^{1,0} \mathcal{L}$.\\
\textbf{Hyper-parameters:}  Grid number $H$, iteration number on $\Omega^{h}: T_{h}$. 

\For{$ h = 1 : H$}{
$\hat{u}_{i,n}^{h,T_{h}}=$Feature Extractor$(\hat{F}_{i,n}^{h}, \hat{u}_{i,n}^{h,0}, T_{h})$.
$\hat{v}_{i,n}^{h,T_{h}} = softmax\big(\frac{-\frac{1}{2}(\hat{f}_{i}^{h}-\hat{u}_{i,n}^{h,T_{h}})^{2}-\frac{\lambda}{2} |\mathcal{L}\hat{u}_{i,n}^{h,T_{h}}|^{2}-\alpha k*(1-2\hat{v}_{i,n}^{h,0})}{\varepsilon}\big)$.\\
\If{$h<H$}{
Compute residual: $\hat{r}_{h}=\hat{F}_{i,n}^{h}-A^{h}\hat{u}_{i,n}^{h,T_{h}}$. 
Compute initial value on $\Omega^{h+1}$: $\hat{u}_{i,n}^{h+1,0}=I_{h}^{h+1}\hat{u}_{i,n}^{h,T_{h}}$, $\hat{v}_{i,n}^{h+1,0}=I_{h}^{h+1}\hat{v}_{i,n}^{h,T_{h}}$. 
$A^{h+1}=\hat{v}_{i,n}^{h+1,0}I+\lambda\mathcal{L}^{*}\hat{v}_{i,n}^{h+1,0}\mathcal{L}$.
$\hat{F}_{i,n}^{h+1} = I_{h}^{h+1}\hat{r}_{h}+A^{h+1}\hat{u}_{i,n}^{h+1,0}$, $\hat{f}_{i}^{h+1}= I_{h}^{h+1}\hat{f}_{i}^{h}.$
}
}
\For{$ h = H : 2$}{
Prolongation: $\hat{u}_{i,n}^{h-1,T_{h-1}}=\hat{u}_{i,n}^{h-1,T_{h-1}}+I_{h}^{h-1}(\hat{u}_{i,n}^{h,T_{h}}-\hat{u}_{i,n}^{h,0})$, $\hat{v}_{i,n}^{h-1,T_{h-1}} =I_{h}^{h-1}\hat{v}_{i,n}^{h,T_{h}}$. 
Error smooth: $\hat{u}_{i,n}^{h-1,T_{h-1}} = $Feature Extractor$(\hat{F}_{i,n}^{h-1}, \hat{u}_{i,n}^{h-1,T_{h-1}}, T_{h-1})$.\\
$\hat{v}_{i,n}^{h-1,T_{h-1}}= softmax\big(\frac{-\frac{1}{2}(\hat{f}_{i}^{h-1}- \hat{u}_{i,n}^{h-1,T_{h-1}})^{2}-\lambda|\mathcal{L}\hat{u}_{i,n}^{h-1,T_{h-1}}|^{2} -\alpha k*(1-2\hat{v}_{i,n}^{h-1,T_{h-1}})}{\varepsilon}\big)   $.
}
\textbf{Output:} $\hat{u}_{i,n} =\hat{u}_{i,n}^{1,T_{1}}$, $\hat{v}_{i,n}=\hat{v}_{i,n}^{1,T_{1}}$.
\end{algorithm}

\section{MSNet for image segmentation}\label{net}
We will design a segmentation network using the unrolling technique according to Algorithm \ref{algorithm4.1} and Algorithm \ref{algorithm4.2}. 

We start by using a convolution layer with a kernel size of $3\times3$ as the head network to map the input images to a high-dimensional feature domain, i.e. $\hat{f} = \mathcal{G}(f)$. Let $\hat{\mathbf{u}}^{0}=\hat{\mathbf{f}}=(\hat{f},\hat{f},...,\hat{f})\in\mathbb{R}^{IN\times H\times W}$, where H and W respectively represent the height and width of the image. Then, we feed $\hat{\mathbf{u}}^{0}$ into the intermediate layers to obtain the updated $\hat{\mathbf{u}}^{t}$ and new $\hat{\mathbf{v}}^{t}$ for $t=1,2,..., T$. Finally, the output of the segmentation network is $\mathbf{v}\in\mathbb{R}^{N\times H\times W}$, which is obtained by applying $\bar{\mathcal{G}}$, a combination of two convolutions layers with a kernel size of $3\times3$.

Now, we utilize the unrolling technique to construct the intermediate layers of the network based on the updates of $\hat{\mathbf{u}}$ and $\hat{\mathbf{v}}$ in \ref{algorithm4.1}. According to Algorithm \ref{algorithm4.1}, we obtain:
\begin{equation}\label{equ5.1}
\hat{\mathbf{u}}^{t_{1}+1} = \hat{\mathbf{u}}^{t_{1}} + \triangle t\big(\hat{\mathbf{f}}\hat{\mathbf{v}}^{t_{1}}-\mathcal{A}\hat{\mathbf{u}}^{t_{1}}\big),
\end{equation}
where $\mathcal{A}=\hat{\mathbf{v}}^{t_{1}}+\lambda \mathcal{L}^{*}\hat{\mathbf{v}}^{t_{1}}\mathcal{L}.$
\begin{equation}\label{equ5.2}
\hat{\mathbf{v}}^{t_{1}+1}= softmax\big(\frac{-\frac{1}{2}(\hat{\mathbf{f}}-\hat{\mathbf{u}}^{t_{1}+1})^{2}-\frac{\lambda}{2}|\mathcal{L}\hat{\mathbf{u}}^{t_{1}+1}|^{2}-\alpha k*(1-2\hat{\mathbf{v}}^{t_{1}})}{\varepsilon}\big),
\end{equation}
where $\hat{\mathbf{u}}, \hat{\mathbf{f}}\in\mathbb{R}^{C\times H\times W}$, $C=IN$, $|\Omega|=HW$.

\begin{figure}[htbp]
\centering
\includegraphics[width=0.6\linewidth]{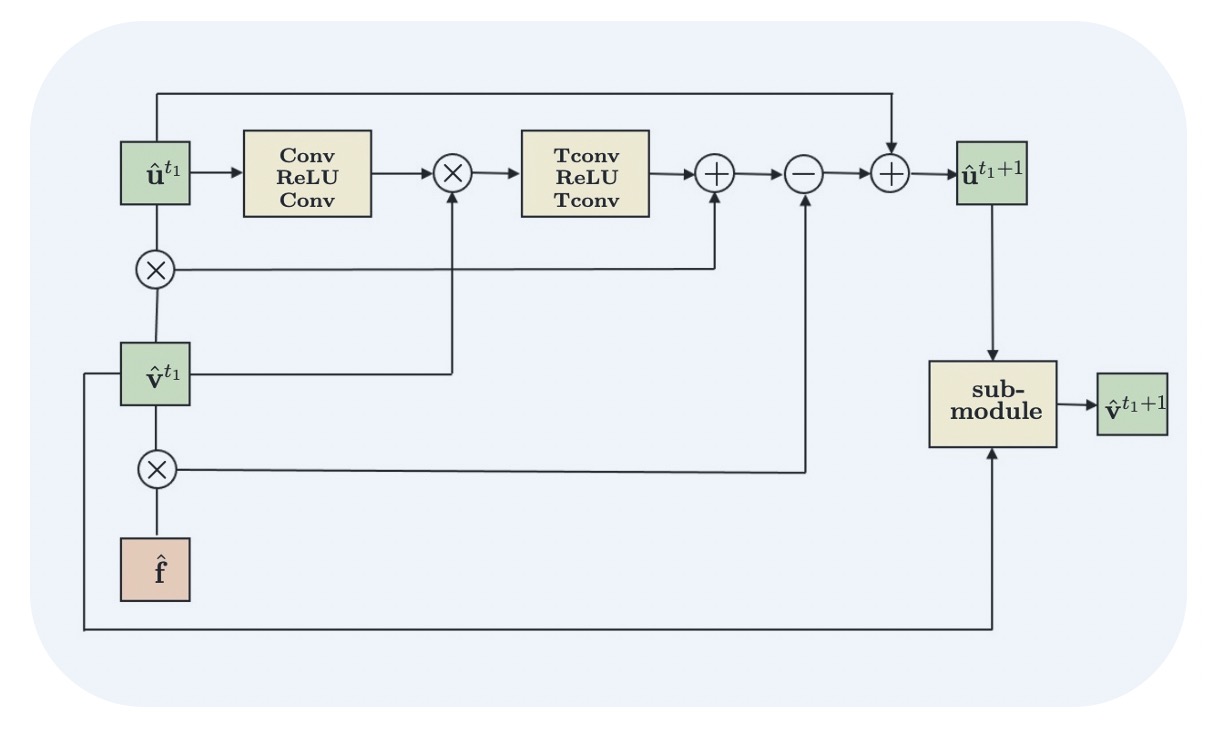}
\caption{Basic feature extractor module (FEM).}\label{fig1}
\end{figure}

\begin{figure}[htbp]
\centering
\includegraphics[width=0.6\linewidth]{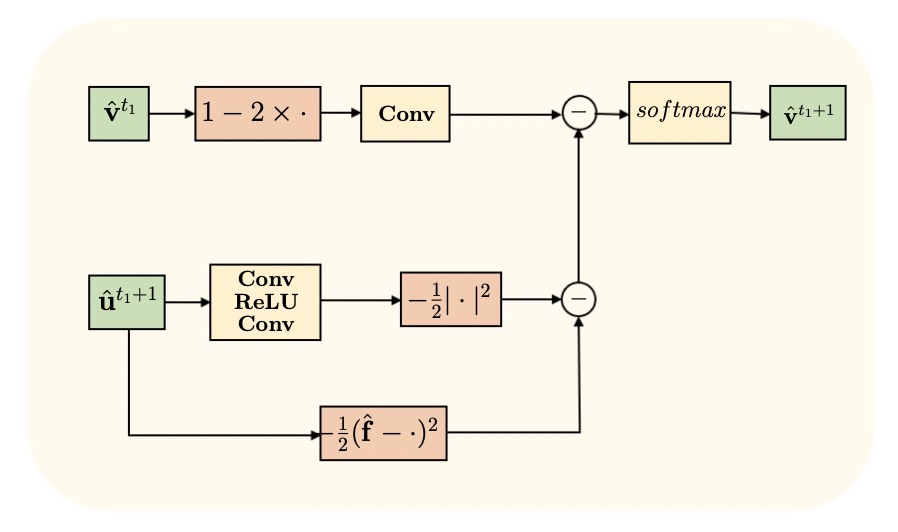}
\caption{Submodule.}\label{submodule}
\end{figure}

We first design a fundamental feature extractor module (FEM) based on the equations (\ref{equ5.1}) and (\ref{equ5.2}). 
In our proposed generalized MS model, $\mathcal{L}$ represents a learnable operator. We define $\mathcal{L}$ as a learnable operator that comprises of convolution, ReLU activation, and another convolution. Additionally, we define $\mathcal{L}^{*}$ as the corresponding transpose convolution, ReLU activation, and another transpose convolution. These operators serve as the essential components of our base feature extraction module.
In \figurename~\ref{fig1}, we illustrate the basic FEM by unrolling techniques for the equations (\ref{equ5.1}) and (\ref{equ5.2}). 
Specifically, the submodule shown in \figurename~\ref{fig1} is unrolled by equation (\ref{equ5.2}), which updates $\hat{\mathbf{v}}^{t_{1}+1}$ based on the updated $\hat{\mathbf{u}}^{t_{1}+1}$ and the original $\hat{\mathbf{v}}^{t_{1}}$, where $k$ is learned through convolution operator, and parameters are absorbed by the subsequent learnable operator. This weight learning submodule guides the feature extraction process, which is shown in \figurename~\ref{submodule}. 

Until now, we have constructed the intermediate layers of the network by stacking basic feature extractor modules (FEMs).

Now, we summarize the overall architecture of the network. In the first layer, the images are processed through a convolution layer to obtain high-dimensional features. These features are then passed into the Feature Extractor Modules (FEMs), which include weight learning submodules. 
Finally, we use convolutional layers to obtain the network's output. However, the network extract features at a single scale. We have tried using this network for segmentation, but the results were particularly poor. 
In order to enhance the accuracy and efficiency of image segmentation, we unfold Algorithm \ref{algorithm4.2} obtained through the multi-grid algorithm to create a network for multi-scale feature extraction.

\subsection{The multi-grid framework for building the U-type MSNet}
Many successful and widely used network structures in deep learning are based on inspiration from classic algorithms. In \cite{he2019mgnet}, the authors established MgNet by studying the connections between convolutional neural networks and the traditional multi-grid method. MgNet provides a unified framework that connects multi-grid and CNN. For example, the U-type network architecture can be constructed by unrolling the multi-grid algorithm for solving partial differential equations (PDEs). Motivated by this idea, we will design a U-type network architecture for our method based on the multi-grid algorithm presented in Algorithm \ref{algorithm4.2}.
 
\begin{figure}[htbp]
\centering
\includegraphics[width=0.9\linewidth]{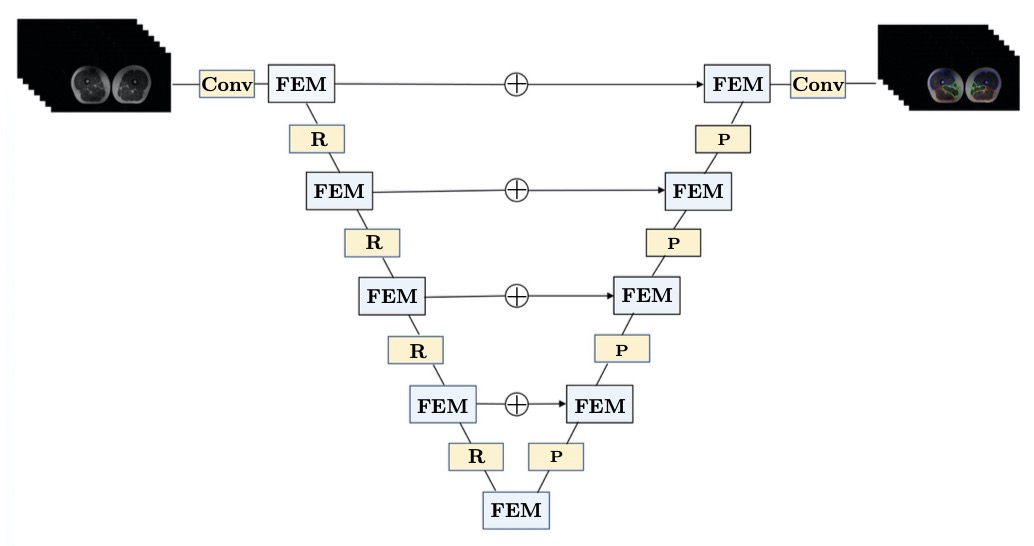}
\caption{MSNet. R: restriction operator, P: prolongation operator.}\label{fig2}
\end{figure}

The multi-grid algorithm relies on two important components, the restriction operator and prolongation operator, to transfer information between grids. Typically, pooling and interpolation with fixed parameters are used to implement these operators. In our network, we implement the restriction operator using a $3\times 3$ convolution with a stride of 2, and we use interpolation for prolongation. To reduce computation cost, we set the relaxation iteration number to $T_{h}=1$ in Algorithm \ref{algorithm4.2}. By combining the basic feature extractor module (FEM) and setting the grid number $H=5$, we construct the U-type segmentation network structure using the unrolling technique. The resulting segmentation network, named MSNet, is illustrated in \figurename~\ref{fig2}. By replacing all 2D operations in the proposed network with 3D operations, the network can also be used for 3D image segmentation.

By utilizing multi-grid techniques, our network achieves multi-scale feature extraction similar to U-Net. However, unlike the black-box nature of U-Net, the feature extraction process of our network at each scale is guided by a variational model-based algorithm, which makes the feature extraction process mathematically interpretable.

\section{Experimental results}\label{experiment}
 \subsection{Implementation details}
 The framework is implemented in Python based on Pytorch. We use Adam optimizer for training the network. Initial learning rate is $1\times10^{-3}$ and it is multiplied by 0.8 every 200 iterations. The feature channels in the proposed network are set to $C=16$.
 In order to compare with other deep learning methods, we only modified the network structure while keeping other settings such as data processing methods and loss function selection unchanged. This ensures that we are comparing the performance differences of different network structures, rather than the impact of other setting differences.
 We choose a soft multi-class Dice loss \cite{xu2019deepatlas} as segmentation loss function, which addresses imbalances inherently:
 
 $$DiceLoss(y, y^{*}) = 1 - \frac{1}{N} \sum_{n=1}^{N} \frac{2 \sum_{j=1}^{J} y_{jn} y_{jn}^{*} + \epsilon}{\sum_{j=1}^{J} y_{jn} + \sum_{j=1}^{J} y_{jn}^{*} + \epsilon}$$
where $y_{jn}^{*}$ denotes the ground truth label for pixel $j$ in class $n$, and $y_{jn}$ denotes the predicted label for pixel $j$ in class $n$, $N$ is the number of classes, $J$ is the total number of pixels, and $\epsilon$ is a small constant added to the denominator to avoid division by zero.
 \subsection{Evaluation Metrics}
 We will use Accuracy (Acc), intersection over union (IoU), Dice similarity coefficient (DSC), and average surface distance (ASD) as evaluation metrics for our algorithm's segmentation results. Let X represent the segmentation result and Y represent the ground truth. Additionally, let $\partial X$ and $\partial Y$ denote the segmentation boundary of X and the boundary of Y, respectively. We define these terms as follows:
 
 Accuracy:
 $$Acc=\frac{TP + TN}{TP + FP + TN + FN},$$
 where TP is the number of true positive, FP represents the number of false positive, FN is the number of false negative, and TN is the number of false negative.
 
 IoU:
 $$IoU = \frac{|X \cap Y| }{ |X \cup Y|}.$$
 
 Dice Similarity Coefficient:
 $$DSC(X,Y)=\frac{2|X\cap Y|}{|X|+|Y|}.$$
 
 Average Surface Distance:
 $$ASD(\partial X,\partial Y) = \frac{\sum\limits_{x\in\partial X}\min\limits_{y\in\partial Y}\|x-y\|+\sum\limits_{y\in\partial Y}\min\limits_{x\in\partial X}\|y-x\|}{|\partial X|+|\partial Y|}.$$

In addition, for multi-class segmentation, the following metrics can be used for evaluation:
  $Acc_{avg} = \frac{1}{N}\sum_{n=1}^{N}Acc_{n},$
  $mIoU = \frac{1}{N}\sum_{n=1}^{N}IoU_{n},$
 $DSC_{avg} = \frac{1}{N}\sum_{n=1}^{N}DSC_{n},$
 $ASD_{avg} = \frac{1}{N}\sum_{n=1}^{N}ASD_{n}.$
 Where $N$ represents the total number of classes, $IoU_{n}$ is the IoU for class $n$, $DSC_{n}$ is the DSC for class $n$, $Acc_{n}$ is the Acc for class $n$, and $ASD_{n}$ is the ASD for class $n$.

\subsection{Experiments on  REFUGE dataset}
Accurately segmenting the optic disc (OD) and optic cup (OC) from fundus images is crucial for effective screening and diagnosis of glaucoma. In this experiment, we choose REFUGE dataset \cite{orlando2020refuge} from the public optic disc (OD) and optic cup (OC) datasets. We cropped and resized the images to a pixel resolution of $128\times 128$ for our experiments. To evaluate the performance of our method, we conducted OD and OC segmentation experiments using different training set sizes: 20, 50, 100, and 400 images, with a corresponding test set of 400 images for each. The performance metrics, including the average values of $Acc_{avg}$, mIoU, $DSC_{avg}$, and $ASD_{avg}$ for 400 test images are presented in Table~\ref{table2}. 

\begin{figure}[htbp]
  \centering
  \hfill
  \begin{subfigure}[t]{0.22\textwidth}
    \centering
    \includegraphics[width=\textwidth]{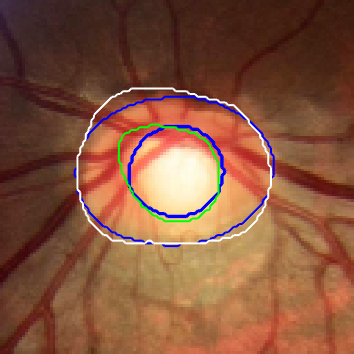}
  \end{subfigure}%
  \hfill
  \begin{subfigure}[t]{0.22\textwidth}
    \centering
    \includegraphics[width=\textwidth]{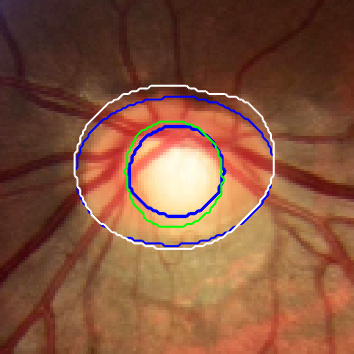}
  \end{subfigure}%
  \hfill
  \begin{subfigure}[t]{0.22\textwidth}
    \centering
    \includegraphics[width=\textwidth]{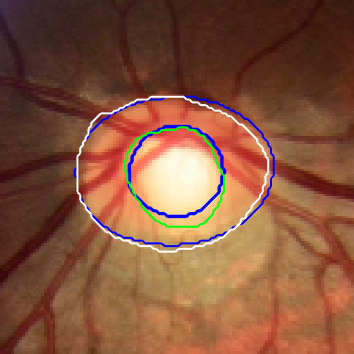}
  \end{subfigure}
  \hfill 
  \begin{subfigure}[t]{0.22\textwidth}
    \centering
    \includegraphics[width=\textwidth]{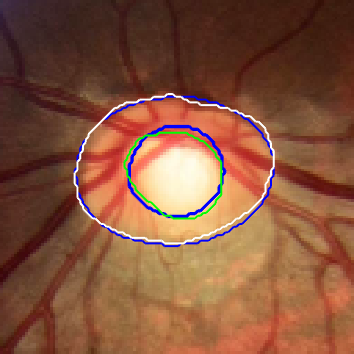}
  \end{subfigure}  \\
  \hfill
  \begin{subfigure}[t]{0.22\textwidth}
    \centering
    \includegraphics[width=\textwidth]{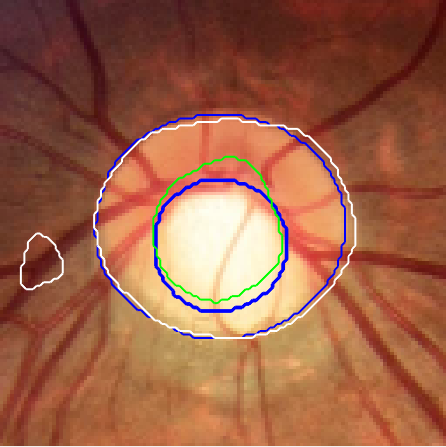}
  \end{subfigure}%
  \hfill
  \begin{subfigure}[t]{0.22\textwidth}
    \centering
    \includegraphics[width=\textwidth]{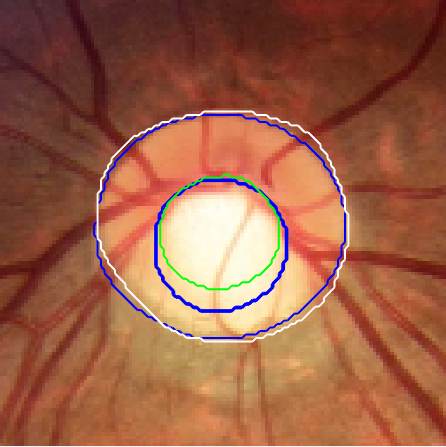}
  \end{subfigure}
  \hfill 
  \begin{subfigure}[t]{0.22\textwidth}
    \centering
    \includegraphics[width=\textwidth]{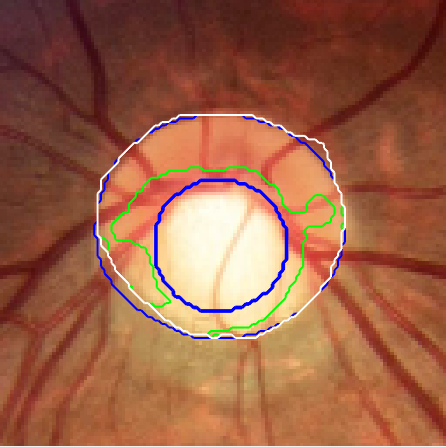}
  \end{subfigure}
\hfill
  \begin{subfigure}[t]{0.22\textwidth}
    \centering
    \includegraphics[width=\textwidth]{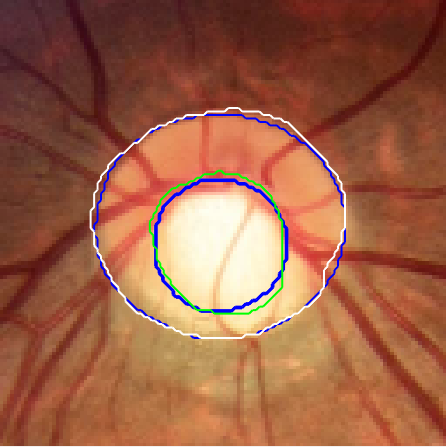}
  \end{subfigure} \\
\hfill
  \begin{subfigure}[t]{0.22\textwidth}
    \centering
    \includegraphics[width=\textwidth]{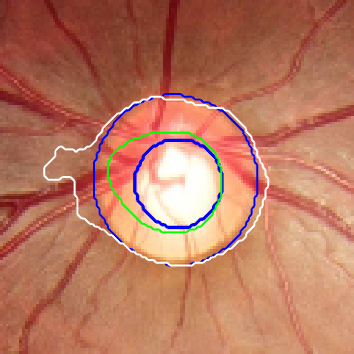}
  \end{subfigure} 
  \hfill
  \begin{subfigure}[t]{0.22\textwidth}
    \centering
    \includegraphics[width=\textwidth]{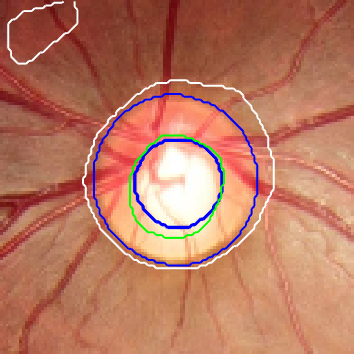}
  \end{subfigure}
\hfill
  \begin{subfigure}[t]{0.22\textwidth}
    \centering
    \includegraphics[width=\textwidth]{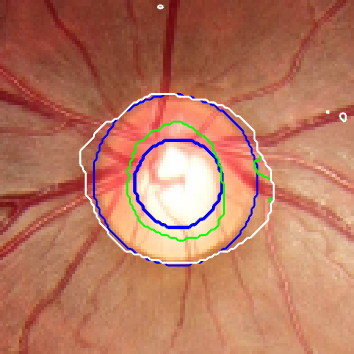}
  \end{subfigure}
  \hfill
  \begin{subfigure}[t]{0.22\textwidth}
    \centering
    \includegraphics[width=\textwidth]{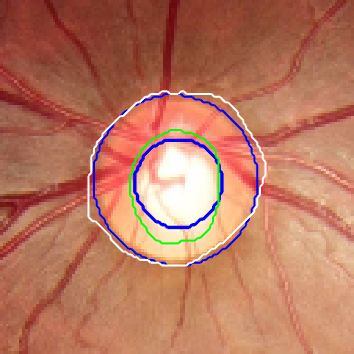}
  \end{subfigure}\\
  \hfill
  \begin{subfigure}[t]{0.22\textwidth}
    \centering
    \includegraphics[width=\textwidth]{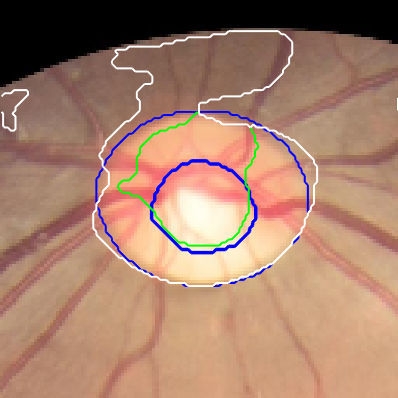}
  \end{subfigure} 
  \hfill
  \begin{subfigure}[t]{0.22\textwidth}
    \centering
    \includegraphics[width=\textwidth]{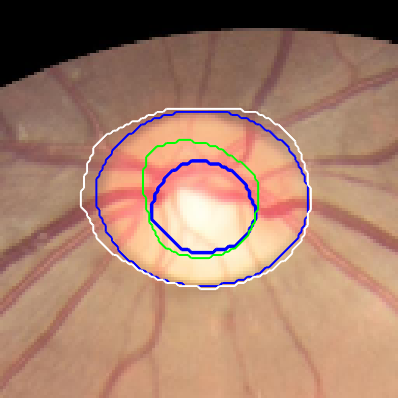}
  \end{subfigure}
\hfill
  \begin{subfigure}[t]{0.22\textwidth}
    \centering
    \includegraphics[width=\textwidth]{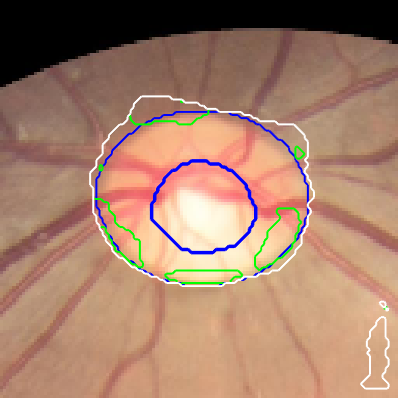}
  \end{subfigure}
  \hfill
  \begin{subfigure}[t]{0.22\textwidth}
    \centering
    \includegraphics[width=\textwidth]{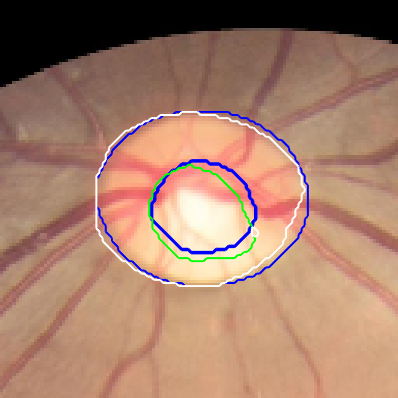}
  \end{subfigure}\\
  \hfill
  \begin{subfigure}[t]{0.22\textwidth}
    \centering
    \includegraphics[width=\textwidth]{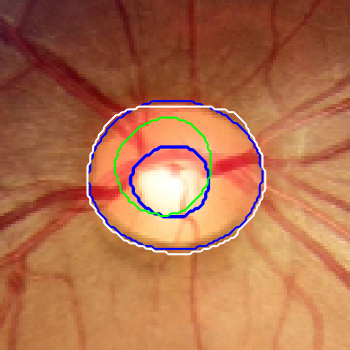}
    \caption{U-Net \cite{ronneberger2015u}}
  \end{subfigure} 
  \hfill
  \begin{subfigure}[t]{0.22\textwidth}
    \centering
    \includegraphics[width=\textwidth]{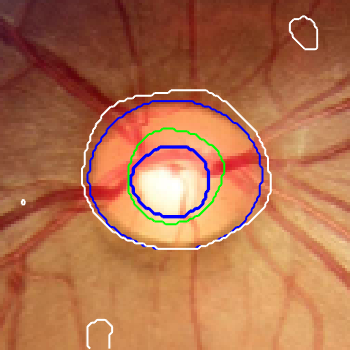}
    \caption{DeepLabV3+ \cite{chen2018encoder}}
  \end{subfigure}
\hfill
  \begin{subfigure}[t]{0.22\textwidth}
    \centering
    \includegraphics[width=\textwidth]{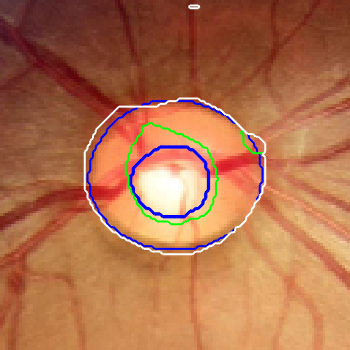}
    \caption{R2U-Net \cite{alom2018recurrent}}
  \end{subfigure}
  \hfill
  \begin{subfigure}[t]{0.22\textwidth}
    \centering
    \includegraphics[width=\textwidth]{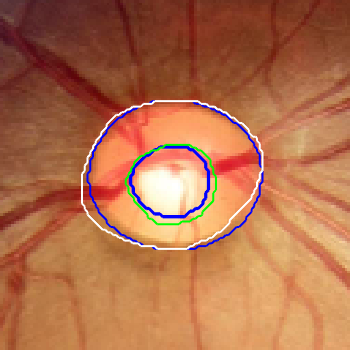}
    \caption{Proposed}
  \end{subfigure}\\
  \caption{Example segmentation results of the different methods on REFUGE dataset. White lines and green lines indicate predicted results, and blue lines indicate ground truths.} \label{fig4}
  \end{figure}

By analyzing the data presented in the table Table~\ref{table2}, it can be concluded that the proposed method exhibits superior performance when compared to the other methods. This can be observed in terms of Accuracy, mIoU, and DSC scores, where the proposed method achieves the highest scores, indicating better accuracy and segmentation results. Moreover, the proposed method also has the smallest ASD score, further highlighting its superior performance compared to the other methods. To provide visual evidence of the segmentation quality, we present a visual comparison of the optic disc and cup segmentation results in \figurename~\ref{fig4} when the training data consists of 400 images. Through the visualization of segmentation boundaries in \figurename~\ref{fig4}, we have found that our segmentation lines closely resemble the ground truth labels, indicating improved alignment and matching. Furthermore, our segmentation results exhibit smoother and more continuous boundaries, reducing the occurrence of interruptions or fractures in the segmented regions. Overall, our method enhances the accuracy of the segmented regions, improving the overall segmentation performance.

\begin{table}[htbp]
\small
\caption{Results of U-Net, DeepLabv3+, R2U-Net, and the proposed network on REFUGE dataset.}\label{table2}
\begin{center}
\setlength{\tabcolsep}{4.8mm}{
  \begin{tabular}{c cc ccc } 
\hline
Data & Metrics & U-Net\cite{ronneberger2015u}  & DeepLabV3+\cite{chen2018encoder}   & R2U-Net\cite{alom2018recurrent} & Proposed \\
\hline
\multirow{4}{*}{S=20}&$Acc_{avg}$&0.9028& 0.9165     & 0.9320&  \textbf{0.9495}\\
&mIoU& 0.6236& 0.6319 & 0.6105& \textbf{0.7118}\\
&$DSC_{avg}$&  0.7416 &0.7492 & 0.7192& \textbf{0.8147}\\
&$ASD_{avg}$ & 3.3287 &2.0687 &3.7283 & \textbf{2.0031}\\
\hline
\multirow{4}{*}{S=50}&$Acc_{avg}$&0.8701& 0.9269     & 0.9205&  \textbf{0.9486}\\
&mIoU& 0.5488& 0.6499 & 0.6551& \textbf{0.6855}\\
&$DSC_{avg}$&  0.6665 &0.7649 & 0.7698& \textbf{0.7904}\\
&$ASD_{avg}$ & 5.2768 &1.7757 &3.3055 & \textbf{1.6800}\\
\hline
\multirow{4}{*}{S=100}&$Acc_{avg}$&0.9071& 0.9194     & 0.9314&  \textbf{0.9553}\\
&mIoU& 0.6129& 0.6568 & 0.6672& \textbf{0.7497}\\
&$DSC_{avg}$&  0.7321 &0.7710 & 0.7785& \textbf{0.8448}\\
&$ASD_{avg}$ & 0.7441 &2.3399 &3.0226 & \textbf{0.9986}\\
\hline
\multirow{4}{*}{S=400}&$Acc_{avg}$&0.9600& 0.9465     & 0.9471&  \textbf{0.9649 }\\
&mIoU& 0.7269& 0.6984 & 0.6436& \textbf{0.7597 }\\
&$DSC_{avg}$&  0.8233 &0.8020 & 0.7492& \textbf{0.8520 }\\
&$ASD_{avg}$ & 1.3527 &1.3629 &2.2458 & \textbf{0.8626 }\\
\hline
\end{tabular}}
\end{center}
\end{table}

 \subsection{Experiments on WBC dataset}
 
\begin{figure}[htbp]
  \centering
  \hfill
  \begin{subfigure}[t]{0.22\textwidth}
    \centering
    \includegraphics[width=\textwidth]{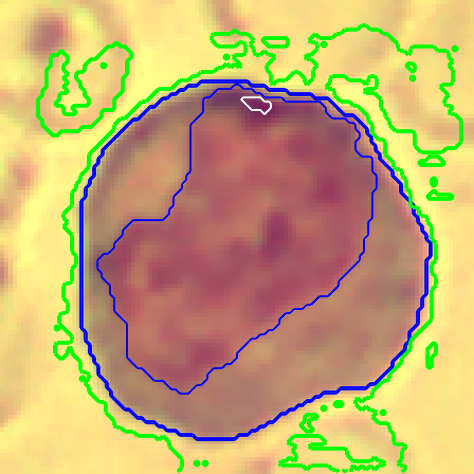}
  \end{subfigure}%
  \hfill
  \begin{subfigure}[t]{0.22\textwidth}
    \centering
    \includegraphics[width=\textwidth]{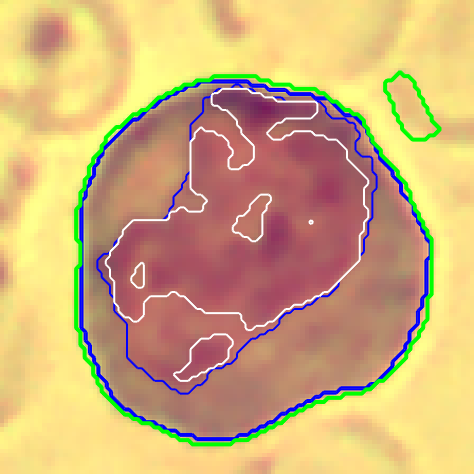}
  \end{subfigure}%
  \hfill
  \begin{subfigure}[t]{0.22\textwidth}
    \centering
    \includegraphics[width=\textwidth]{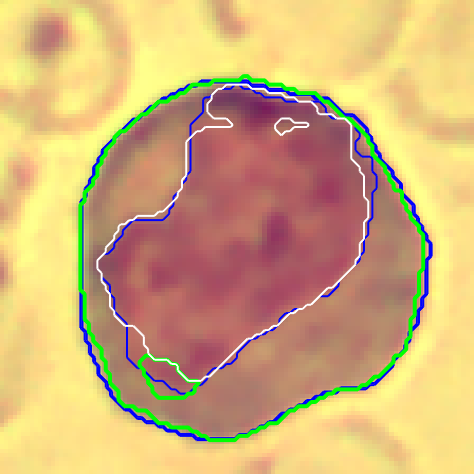}
  \end{subfigure}
  \hfill 
  \begin{subfigure}[t]{0.22\textwidth}
    \centering
    \includegraphics[width=\textwidth]{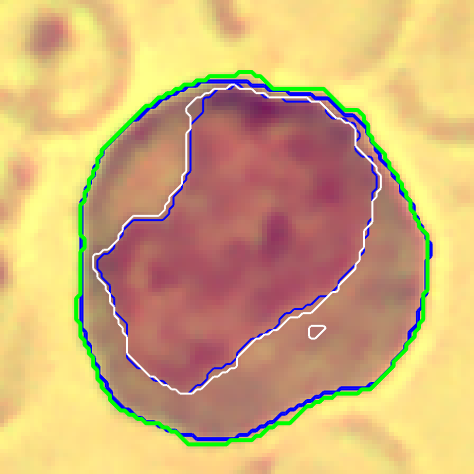}
  \end{subfigure}  \\
  \hfill
  \begin{subfigure}[t]{0.22\textwidth}
    \centering
    \includegraphics[width=\textwidth]{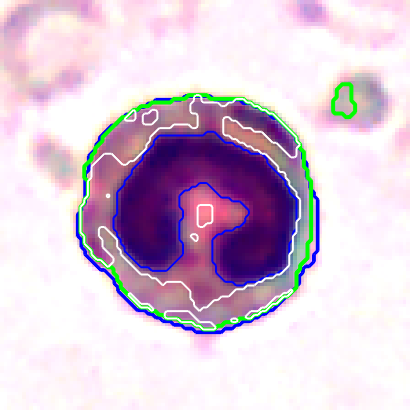}
  \end{subfigure}%
  \hfill
  \begin{subfigure}[t]{0.22\textwidth}
    \centering
    \includegraphics[width=\textwidth]{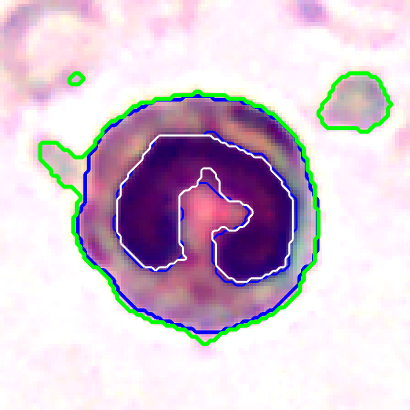}
  \end{subfigure}
  \hfill 
  \begin{subfigure}[t]{0.22\textwidth}
    \centering
    \includegraphics[width=\textwidth]{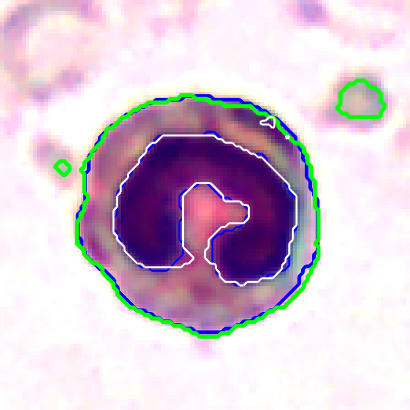}
  \end{subfigure}
\hfill
  \begin{subfigure}[t]{0.22\textwidth}
    \centering
    \includegraphics[width=\textwidth]{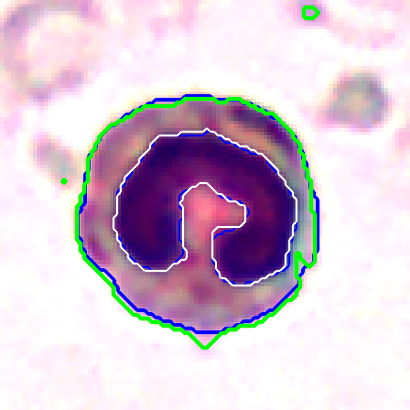}
  \end{subfigure} \\
\hfill
  \begin{subfigure}[t]{0.22\textwidth}
    \centering
    \includegraphics[width=\textwidth]{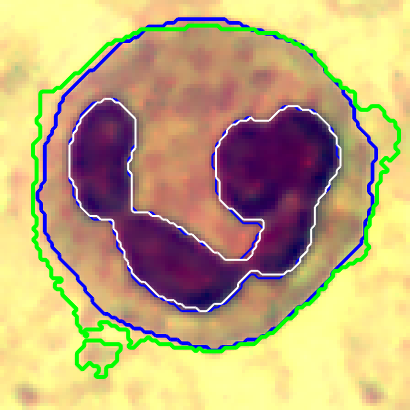}
  \end{subfigure} 
  \hfill
  \begin{subfigure}[t]{0.22\textwidth}
    \centering
    \includegraphics[width=\textwidth]{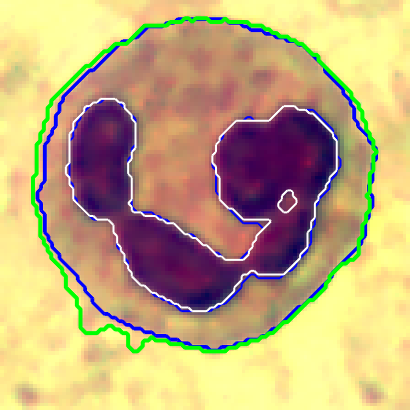}
  \end{subfigure}
\hfill
  \begin{subfigure}[t]{0.22\textwidth}
    \centering
    \includegraphics[width=\textwidth]{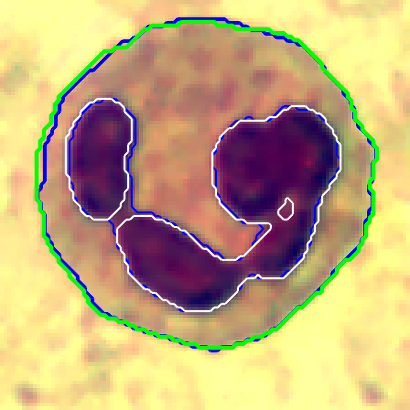}
  \end{subfigure}
  \hfill
  \begin{subfigure}[t]{0.22\textwidth}
    \centering
    \includegraphics[width=\textwidth]{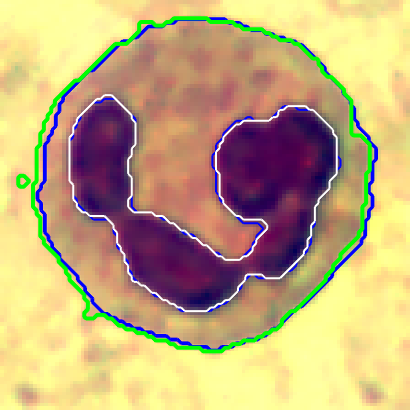}
  \end{subfigure}\\
  \hfill
  \begin{subfigure}[t]{0.22\textwidth}
    \centering
    \includegraphics[width=\textwidth]{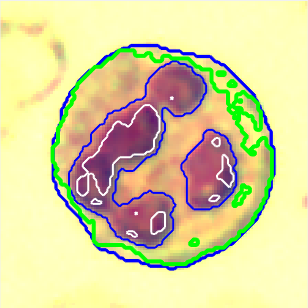}
    \caption{U-Net \cite{ronneberger2015u}}
  \end{subfigure} 
  \hfill
  \begin{subfigure}[t]{0.22\textwidth}
    \centering
    \includegraphics[width=\textwidth]{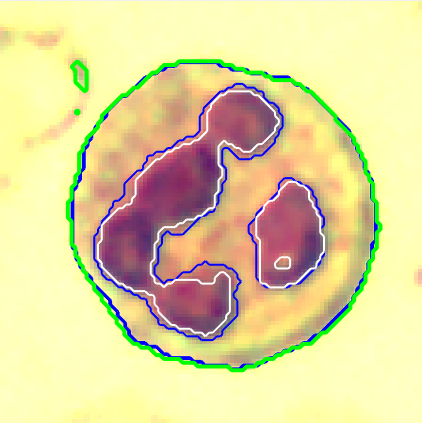}
    \caption{DeepLabV3+ \cite{chen2018encoder}}
  \end{subfigure}
\hfill
  \begin{subfigure}[t]{0.22\textwidth}
    \centering
    \includegraphics[width=\textwidth]{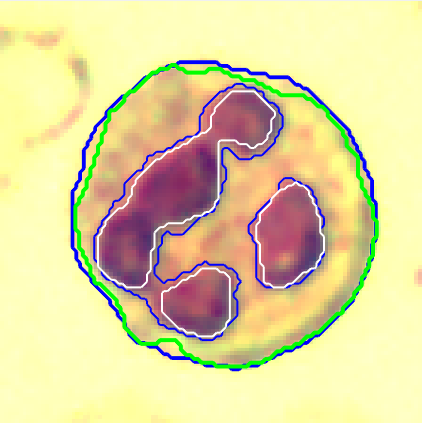}
    \caption{R2U-Net \cite{alom2018recurrent}}
  \end{subfigure}
  \hfill
  \begin{subfigure}[t]{0.22\textwidth}
    \centering
    \includegraphics[width=\textwidth]{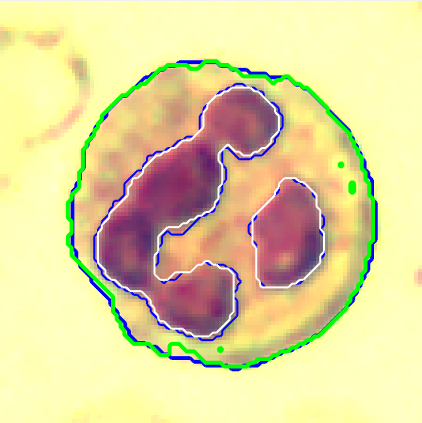}
    \caption{Proposed}
  \end{subfigure}
  \caption{Example segmentation results of the different methods on WBC dataset. White lines and green lines indicate predicted results, and blue lines indicate ground truths.}\label{fig3}
  \end{figure}

 We selected Dataset 1 from White Blood Cell Image Dataset \cite{zheng2018fast} as our experimental dataset, which contains 300 images of size $120\times120$. We randomly selected 10, 30, and 60 images from the dataset as training sets and another 20 images as a test set. All images were cropped to a size of $112\times112$. The average values of $Acc_{avg}$, mIoU, $DSC_{avg}$, and $ASD_{avg}$ for 20 test data are shown in Table~\ref{table1}, which indicates that our proposed method is competitive compared to deep learning networks that lack interpretability.  \figurename~\ref{fig3} shows a visual comparison of the segmentation results for the training data of 30. By observing the segmentation boundaries in \figurename~\ref{fig3}, we can see that the segmentation boundaries of the proposed method exhibit fewer discontinuities and are smoother.

\begin{table}[h]
\small
\caption{Results of U-Net, DeepLabv3+, R2U-Net, and the proposed network on WBC dataset.}\label{table1}
\begin{center}
\setlength{\tabcolsep}{4.8mm}{
  \begin{tabular}{c cc ccc } 
\hline
Data & Metrics & U-Net\cite{ronneberger2015u} & DeepLabV3+\cite{chen2018encoder}  & R2U-Net\cite{alom2018recurrent} & Proposed \\
\hline
\multirow{4}{*}{S=10}&$Acc_{avg}$& 0.9147&0.9492   &  0.9291&  \textbf{0.9509 }\\
&mIoU& 0.7199& 0.8118& 0.7735& \textbf{0.8188 }\\
&$DSC_{avg}$&  0.8056 &0.8856 & 0.8497& \textbf{0.8850 }\\
&$ASD_{avg}$ &2.6567 & \textbf{0.5882} &1.5551&0.7974 \\
\hline
\multirow{4}{*}{S=30}&$Acc_{avg}$& 0.9190  &   0.9772  & 0.9714 &  \textbf{0.9801 }\\
&mIoU&0.7109  & 0.9007 & 0.8895& \textbf{0.9144 }\\
&$DSC_{avg}$&  0.7911  &0.9451 &0.9369 & \textbf{0.9537}\\
&$ASD_{avg}$ &2.3211  &0.2570 &0.5379& \textbf{0.1861}\\
\hline
\multirow{4}{*}{S=60}&$Acc_{avg}$& 0.9395&    0.9829  & \textbf{0.9838} &  0.9819\\
&mIoU&0.7734  &0.9232&\textbf{0.9289} & 0.9215\\
&$DSC_{avg}$&  0.8444 & 0.9589 &\textbf{0.9622} & 0.9579\\
&$ASD_{avg}$ &1.1533  &0.1666&\textbf{0.1469}& 0.1708\\
\hline
\end{tabular}}
\end{center}
\end{table}

  \subsection{Experiments on thigh muscle MR images}
\begin{figure}[ht]
    \centering
    \begin{minipage}[b]{0.3\textwidth}
        \centering
        \includegraphics[width=\textwidth]{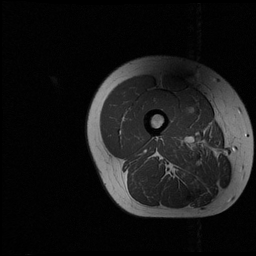}
    \end{minipage}
    \hspace{0.5cm}
    \begin{minipage}[b]{0.3\textwidth}
        \centering
        \includegraphics[width=\textwidth]{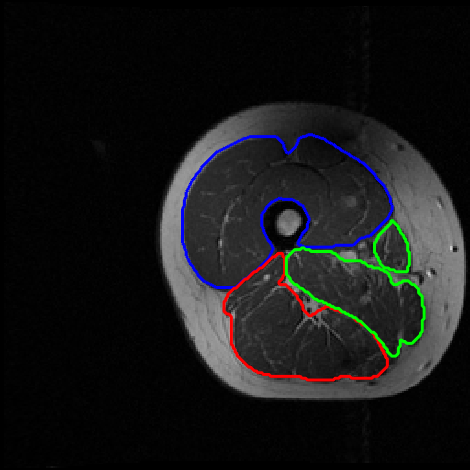}
    \end{minipage}
    \caption{A sample slice of T1-weight thigh muscle MR images. Left: original image. Right: manually segmented objects, with the muscle groups of quadriceps, hamstrings, and others highlighted in blue, red, and green, respectively.}\label{fig8}
    \label{fig:figure_label}
\end{figure}

The segmentation of different thigh muscles is crucial in evaluating musculoskeletal diseases such as osteoarthritis. However, segmentation becomes extremely challenging due to the similarity in intensity and texture between the objects of interest. Moreover, the presence of noise and intensity inhomogeneity further increases the complexity of thigh muscle segmentation. By replacing all the 2D operations in MSNet with 3D operations, we obtain a 3D segmentation network. We evaluate our 3D segmentation network on five T1-weighted MRI 3D volumes, each with a size of $256\times256\times16$. The segmentation of the thigh muscles yielded four distinct clusters, which were identified as follows: quadriceps (vastus lateralis, vastus medialis, rectus femoris, and vastus intermedius), hamstrings (semitendinosus, semimembranosus, biceps femoris short head, and biceps femoris long head), other muscle groups (gracilis, adductor group, and sartorius), and the rest (fat, background, bone marrow, and cortical bone). \figurename~\ref{fig8} illustrates an example of a muscle MRI image.

In the experiment, we tried different traning data sizes and selected 2, 3, and 4 volumes as training sets and used the remaining volumes as corresponding test sets. We compare our proposed network with other three learning-based 3D segmentation networks, including 3D U-Net \cite{cciccek20163d}, MED3D \cite{chen2019med3d}, and DenseVoxNet \cite{yu2017automatic}. To evaluate the performance of 3D U-Net, MED3D, DenseVoxNet and the proposed network, we calculate the average Accuracy, mean Intersection over Union (mIoU), Dice Similarity Coefficient (DSC), and Average Surface Distance (ASD) of the test images. All the results are listed in Table~\ref{table3}. Furthermore, for the training data of 4, the comparison of metrics for Quadriceps, Hamstrings, and Other muscles are shown in Table~\ref{table4}. Based on the data provided in the tables, the results indicate that the proposed method demonstrates superior performance compared to the other methods in terms of Accuracy, mIoU, and DSC scores. Additionally, it achieves the smallest ASD score.

\begin{figure}[htbp]
  \centering
  \begin{subfigure}[t]{0.22\textwidth}
    \centering
    \includegraphics[width=\textwidth]{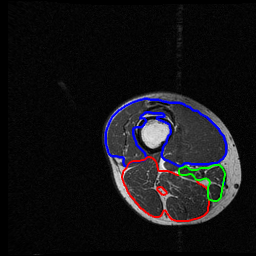}
    \caption{slice=0}
  \end{subfigure}%
  \hfill
  \begin{subfigure}[t]{0.22\textwidth}
    \centering
    \includegraphics[width=\textwidth]{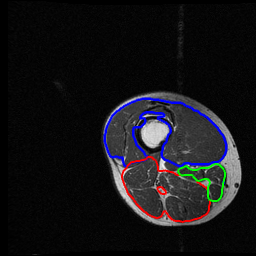}
    \caption{slice=1}
  \end{subfigure}%
  \hfill
  \begin{subfigure}[t]{0.22\textwidth}
    \centering
    \includegraphics[width=\textwidth]{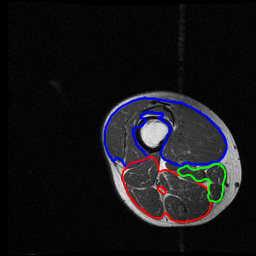}
    \caption{slice=2}
  \end{subfigure}%
  \hfill
  \begin{subfigure}[t]{0.22\textwidth}
    \centering
    \includegraphics[width=\textwidth]{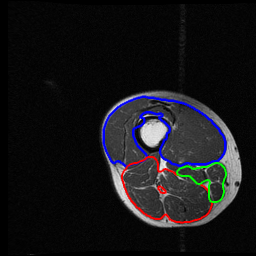}
    \caption{slice=3}
  \end{subfigure}

  \medskip
  
  \begin{subfigure}[t]{0.22\textwidth}
    \centering
    \includegraphics[width=\textwidth]{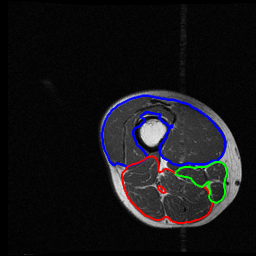}
    \caption{slice=4}
  \end{subfigure}%
  \hfill
  \begin{subfigure}[t]{0.22\textwidth}
    \centering
    \includegraphics[width=\textwidth]{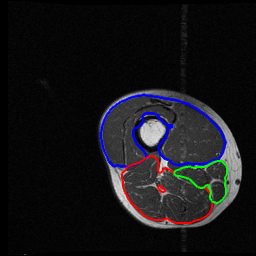}
    \caption{slice=5}
  \end{subfigure}%
  \hfill
  \begin{subfigure}[t]{0.22\textwidth}
    \centering
    \includegraphics[width=\textwidth]{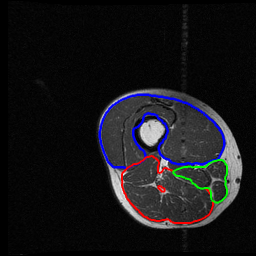}
    \caption{slice=6}
  \end{subfigure}%
  \hfill
  \begin{subfigure}[t]{0.22\textwidth}
    \centering
    \includegraphics[width=\textwidth]{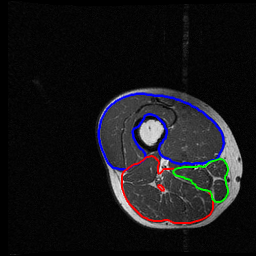}
    \caption{slice=7}
  \end{subfigure}

  \medskip
  
  \begin{subfigure}[t]{0.22\textwidth}
    \centering
    \includegraphics[width=\textwidth]{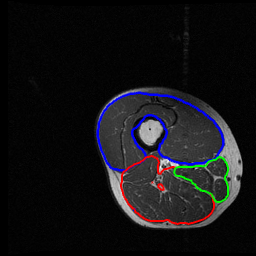}
    \caption{slice=8}
  \end{subfigure}%
  \hfill
  \begin{subfigure}[t]{0.22\textwidth}
    \centering
    \includegraphics[width=\textwidth]{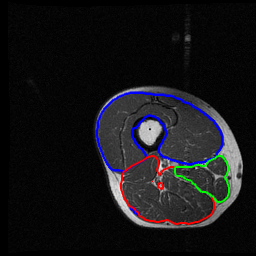}
    \caption{slice=9}
  \end{subfigure}%
  \hfill
  \begin{subfigure}[t]{0.22\textwidth}
    \centering
    \includegraphics[width=\textwidth]{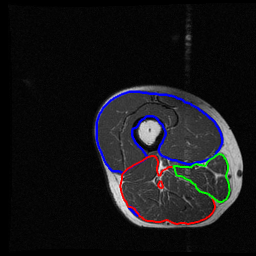}
    \caption{slice=10}
  \end{subfigure}%
  \hfill
    \begin{subfigure}[t]{0.22\textwidth}
    \centering
    \includegraphics[width=\textwidth]{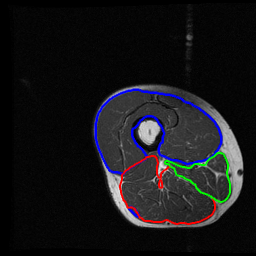}
    \caption{slice=11}
  \end{subfigure}%
  
    \medskip 
    
    \begin{subfigure}[t]{0.22\textwidth}
    \centering
    \includegraphics[width=\textwidth]{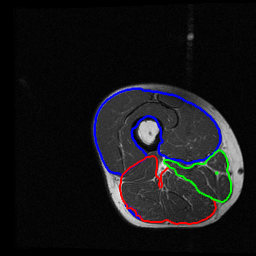}
    \caption{slice=12}
  \end{subfigure}%
  \hfill
    \begin{subfigure}[t]{0.22\textwidth}
    \centering
    \includegraphics[width=\textwidth]{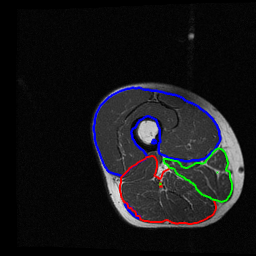}
    \caption{slice=13}
  \end{subfigure}%
  \hfill
    \begin{subfigure}[t]{0.22\textwidth}
    \centering
    \includegraphics[width=\textwidth]{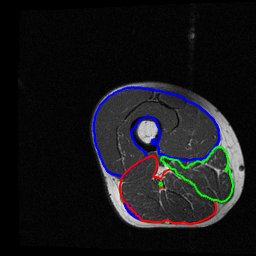}
    \caption{slice=14}
  \end{subfigure}%
    \hfill
    \begin{subfigure}[t]{0.22\textwidth}
    \centering
    \includegraphics[width=\textwidth]{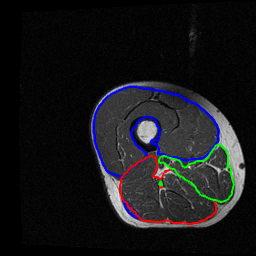}
    \caption{slice=15}
  \end{subfigure}%
  \caption{The segmentation results of all slices by the proposed model with 4 training data.
}\label{fig5}
  \hfill
  \end{figure}

\begin{figure}[htbp]
  \centering
  \begin{subfigure}[t]{0.22\textwidth}
    \centering
    \includegraphics[width=\textwidth]{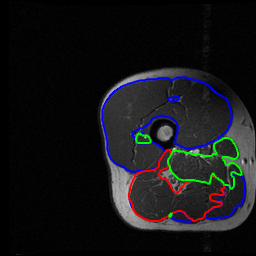}
  \end{subfigure}%
  \hfill
  \begin{subfigure}[t]{0.22\textwidth}
    \centering
    \includegraphics[width=\textwidth]{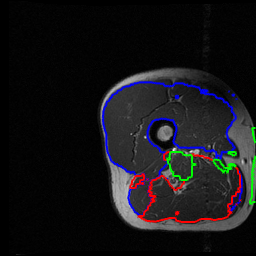}
  \end{subfigure}%
  \hfill
  \begin{subfigure}[t]{0.22\textwidth}
    \centering
    \includegraphics[width=\textwidth]{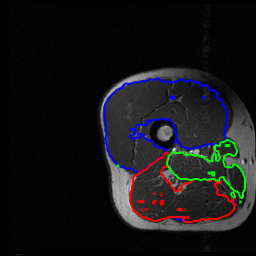}
  \end{subfigure}%
  \hfill
  \begin{subfigure}[t]{0.22\textwidth}
    \centering
    \includegraphics[width=\textwidth]{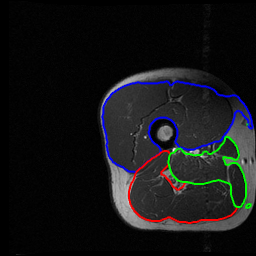}
  \end{subfigure}

  \medskip
  
  \begin{subfigure}[t]{0.22\textwidth}
    \centering
    \includegraphics[width=\textwidth]{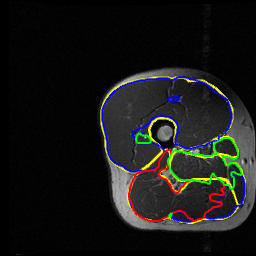}
    \caption{3D U-Net \cite{cciccek20163d}}
  \end{subfigure}%
  \hfill
  \begin{subfigure}[t]{0.22\textwidth}
    \centering
    \includegraphics[width=\textwidth]{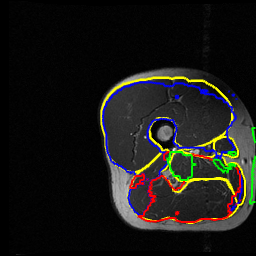}
    \caption{DenseVoxNet \cite{yu2017automatic}}
  \end{subfigure}%
  \hfill
  \begin{subfigure}[t]{0.22\textwidth}
    \centering
    \includegraphics[width=\textwidth]{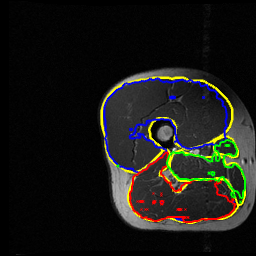}
    \caption{Med3D \cite{chen2019med3d}}
  \end{subfigure}%
  \hfill
  \begin{subfigure}[t]{0.22\textwidth}
    \centering
    \includegraphics[width=\textwidth]{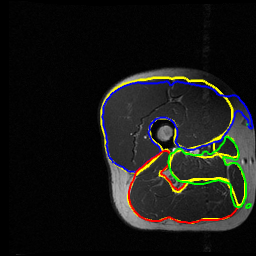}
    \caption{Proposed}
  \end{subfigure}
\caption{Example segmentation results of the different methods on thigh muscle MR images with 3 training data. Red lines, blue lines, and green lines indicate predicted results, and yellow lines indicate ground truths.}\label{fig6}
  \end{figure}

\begin{figure}[htbp]
  \centering
  \begin{subfigure}[t]{0.22\textwidth}
    \centering
    \includegraphics[width=\textwidth]{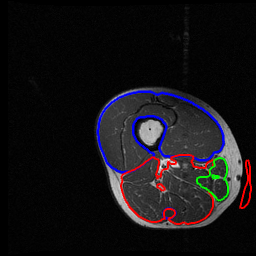}
  \end{subfigure}%
  \hfill
  \begin{subfigure}[t]{0.22\textwidth}
    \centering
    \includegraphics[width=\textwidth]{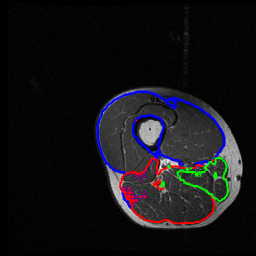}
  \end{subfigure}%
  \hfill
  \begin{subfigure}[t]{0.22\textwidth}
    \centering
    \includegraphics[width=\textwidth]{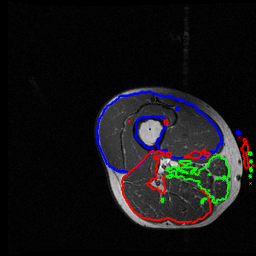}
  \end{subfigure}%
  \hfill
  \begin{subfigure}[t]{0.22\textwidth}
    \centering
    \includegraphics[width=\textwidth]{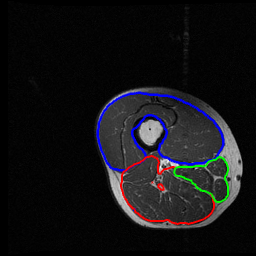}
  \end{subfigure}

  \medskip
  
  \begin{subfigure}[t]{0.22\textwidth}
    \centering
    \includegraphics[width=\textwidth]{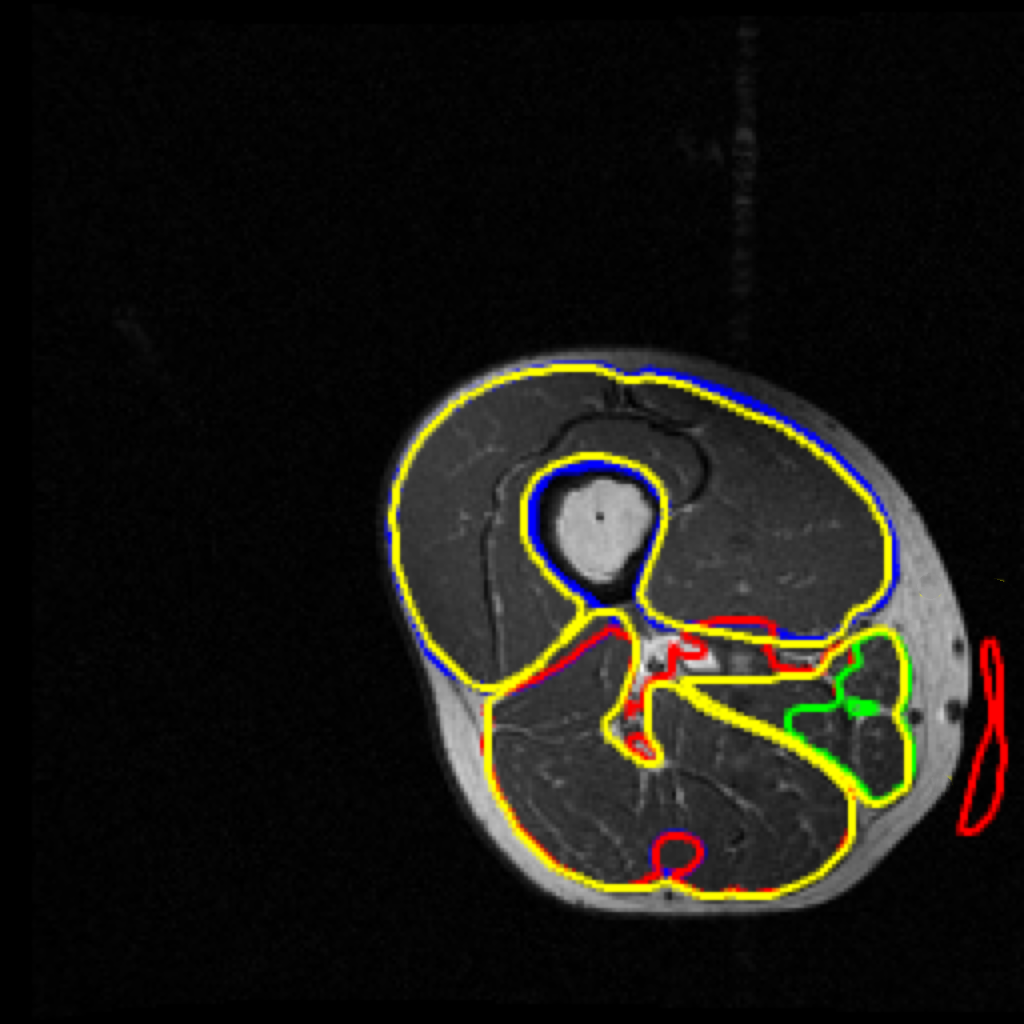}
    \caption{3D U-Net \cite{cciccek20163d}}
  \end{subfigure}%
  \hfill
  \begin{subfigure}[t]{0.22\textwidth}
    \centering
    \includegraphics[width=\textwidth]{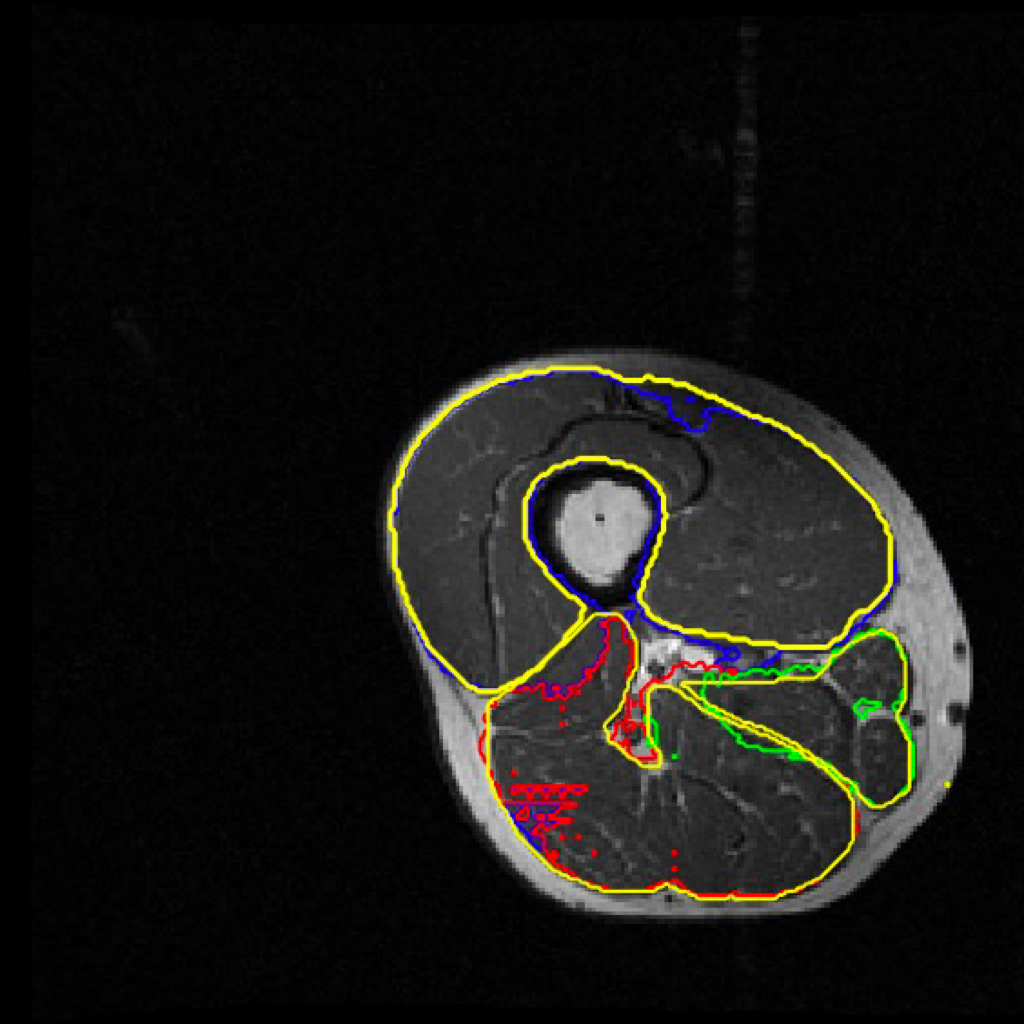}
    \caption{DenseVoxNet \cite{yu2017automatic}}
  \end{subfigure}%
  \hfill
  \begin{subfigure}[t]{0.22\textwidth}
    \centering
    \includegraphics[width=\textwidth]{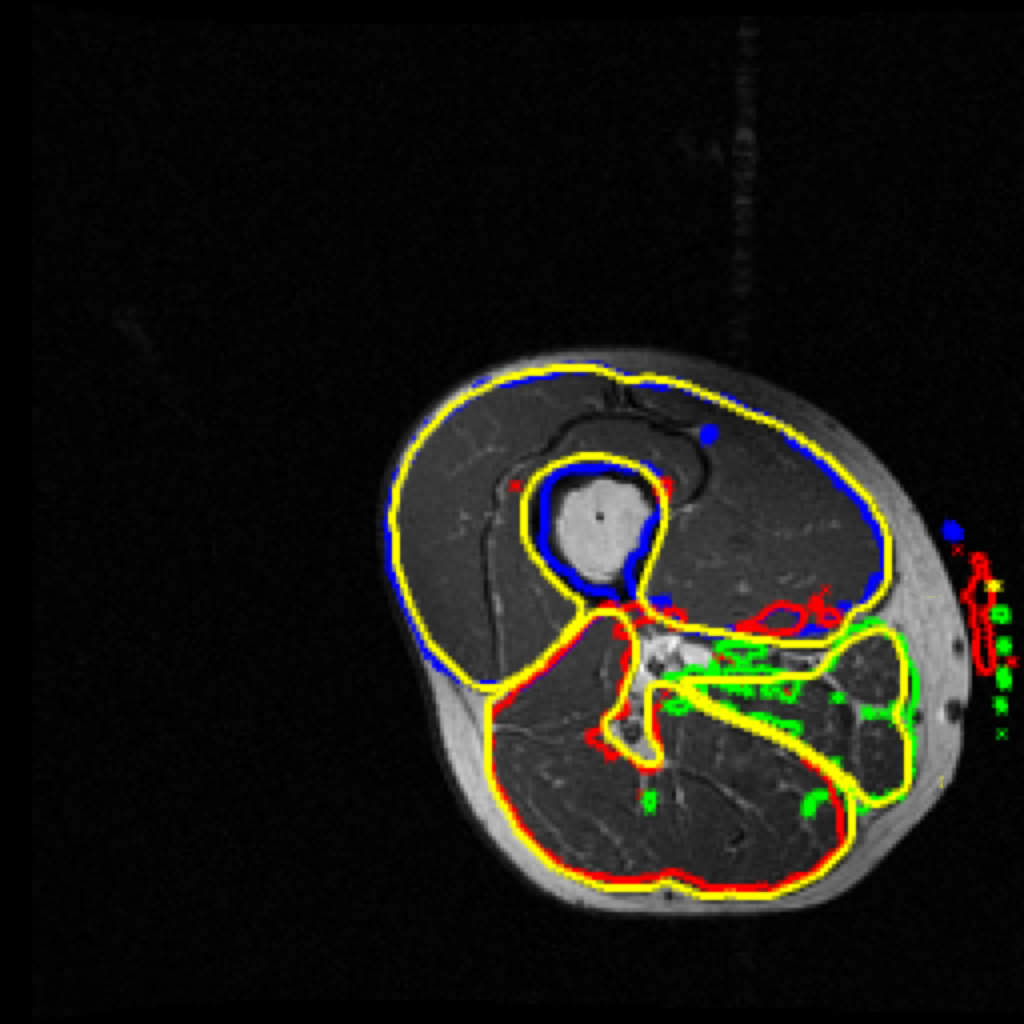}
    \caption{MED3D \cite{chen2019med3d}}
  \end{subfigure}%
  \hfill
  \begin{subfigure}[t]{0.22\textwidth}
    \centering
    \includegraphics[width=\textwidth]{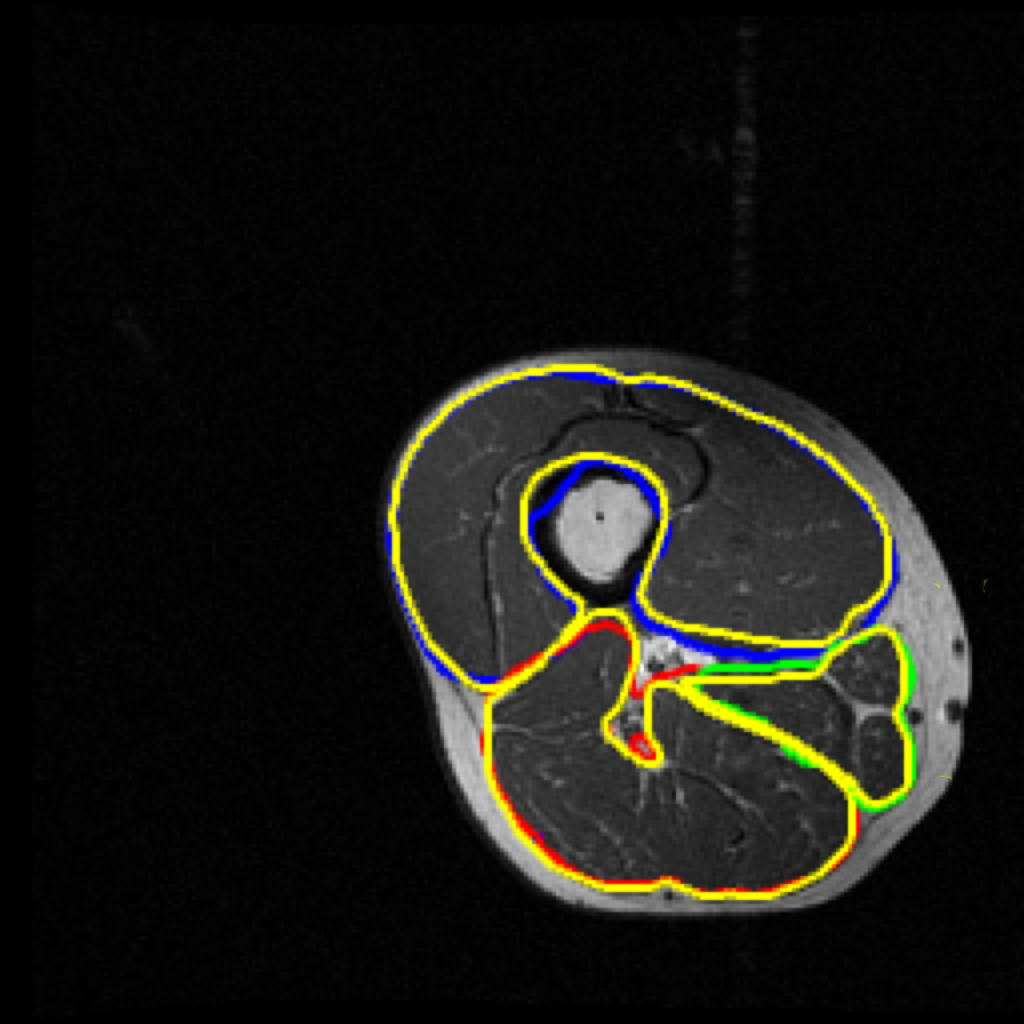}
    \caption{Proposed}
  \end{subfigure}
  \caption{Example segmentation results of the different methods on thigh muscle MR images with 4 training data. Red lines, blue lines, and green lines indicate predicted results, and yellow lines indicate ground truths.}\label{fig7}
  \end{figure}

Experimental results, shown in \figurename~\ref{fig5}, list the segmentation results of the proposed method on all slices with a training set size of 4. It can be observed that the proposed model achieved accurate segmentation for all of the cross sections, indicating its reliability and effectiveness. Now we select the middle axial cross section to visualize the segmentation result. \figurename~\ref{fig6} displays the comparison of the segmentation results obtained with a training set size of 3, while \figurename~\ref{fig7} shows the comparison with a training set size of 4.  As shown in \figurename~\ref{fig6} and \figurename~\ref{fig7}, we find that other methods struggle to differentiate different muscles accurately. In contrast, the proposed method can accurately distinguish each muscle category, avoiding segmentation errors between classes. In addition,
our segmentation boundaries are more consistent with the boundaries of the ground truth labels. Obviously, the segmentation results are closely related to the size of the training data set. Nevertheless, the proposed method exhibits the best performance regardless of the size of the data set.
This indicates that our method is robust and capable of achieving excellent performance even with limited training data. In addition, the experimental results also demonstrate the effectiveness of our proposed network on 3D datasets.

\begin{table}[htbp]
\small
\caption{Results of 3D U-Net, MED3D, DenseVoxNet, and the proposed network on thigh muscle MR image.}\label{table3}
\begin{center}
\setlength{\tabcolsep}{4.8mm}{
  \begin{tabular}{c cc ccc } 
\hline
Data & Metrics &3D U-Net\cite{cciccek20163d} &MED3D\cite{chen2019med3d}  & DenseVoxNet \cite{yu2017automatic} & Proposed \\
\hline
\multirow{4}{*}{S=2}&$Acc_{avg}$&0.8847 & 0.8655     & 0.8704&  \textbf{0.9169}\\
&mIoU& 0.6211 & 0.5607 & 0.5600& \textbf{0.6752}\\
&$DSC_{avg}$&  0.7595 &0.7138 & 0.7097& \textbf{0.7981}\\
&$ASD_{avg}$ & 7.0516 &\textbf{2.5411} &4.1018 & 2.6574\\
\hline
\multirow{4}{*}{S=3} &$Acc_{avg}$&0.9335 & 0.9443  & 0.9110 &  \textbf{0.9711}\\
&mIoU& 0.7371 & 0.7606 &0.6580 & \textbf{0.8023} \\
&$DSC_{avg}$& 0.8451 &0.8609 &0.7851 & \textbf{0.8858}\\
&$ASD_{avg}$ &  4.2504 &0.8812 &1.7728 &\textbf{0.7661}\\
\hline
\multirow{4}{*}{S=4}&$Acc_{avg}$& 0.9565 &  0.9524 & 0.9762& \textbf{0.9862 }\\
&mIoU&0.7538  &0.7286 & 0.8156 & \textbf{0.8519}\\ 
&$DSC_{avg}$& 0.8560 & 0.8377 & 0.8959 & \textbf{0.9185 }\\
&$ASD_{avg}$ &1.7717  & 1.9219  &0.6151& \textbf{0.2899} \\
\hline
\end{tabular}}
\end{center}
\end{table}

\begin{table}[htbp]
\small
\caption{Results of 3D U-Net, MED3D, DenseVoxNet, and the proposed network on thigh muscle MR image with 4 training data.}\label{table4}
\begin{center}
\setlength{\tabcolsep}{3.6mm}{
  \begin{tabular}{c cc ccc } 
\hline
 & Metrics &3D U-Net\cite{cciccek20163d} &MED3D\cite{chen2019med3d}  & DenseVoxNet\cite{yu2017automatic} & Proposed \\
\hline
\multirow{4}{*}{Quadriceps}&Acc&0.9655& 0.9503    & 0.9661 &  \textbf{0.9833}\\
  &IoU& 0.8792& 0.8473 & 0.8811& \textbf{0.8996}\\
&DSC&  0.9357 &0.9174 & 0.9368 &\textbf{0.9471}\\
&ASD&  1.0873 & 0.2755 & \textbf{0.2653} & 0.2799\\
\hline
\multirow{4}{*}{Hamstrings} &Acc&0.9341 & 0.9322  & 0.9758 &  \textbf{0.9801}\\
  &IoU& 0.7466 & 0.7662 &0.8712 & \textbf{0.9001} \\
&DSC& 0.8549 &0.8676 &0.9312 & \textbf{0.9474}\\
&ASD&  2.0296 &3.3472 & 0.3188& \textbf{0.1548}\\
\hline
\multirow{4}{*}{ Other muscles}&Acc& 0.9699 &  0.9746 & 0.9868& \textbf{0.9951 }\\
    &IoU&0.6356  &0.5724 & 0.6945 &  \textbf{0.7560}\\
&DSC& 0.7772 & 0.7281& 0.8197 & \textbf{0.8610 }\\
&ASD&  2.1982 &2.1430 & 1.2612 & \textbf{0.4352}\\
\hline
\end{tabular}}
\end{center}
\end{table}


\subsection{Parameter count comparison}
We calculated the number of parameters for our proposed 2D and 3D segmentation network architectures as well as the other compared network architectures in all the experiments mentioned above. The results are shown in Table~\ref{table5} and Table~\ref{table6}, respectively. Our networks have the smallest parameter count compared to other network architectures, primarily attributed to the utilization of multi-grid techniques. These techniques enable the rapid elimination of parameters, ultimately resulting in our proposed architectures achieving exceptional performance while using fewer parameters.

\begin{table}[h]
\small
\caption{Number of parameters for 2D MSNet.}\label{table5}
\begin{center}
\setlength{\tabcolsep}{4.8mm}{
  \begin{tabular}{c cc ccc } 
\hline
Networks &  U-Net\cite{ronneberger2015u} & DeepLabV3+\cite{chen2018encoder}  & R2U-Net\cite{alom2018recurrent} & Proposed \\
\hline
Nb. of parameters& 31,037,763&5,813,523   & 39,091,523&  \textbf{111,456 }\\
\hline
\end{tabular}}
\end{center}
\end{table}

\begin{table}[h]
\small
\caption{Number of parameters for 3D MSNet.}\label{table6}
\begin{center}
\setlength{\tabcolsep}{4.8mm}{
  \begin{tabular}{c cc ccc } 
\hline
Networks &3D U-Net\cite{cciccek20163d} &MED3D\cite{chen2019med3d}  & DenseVoxNet\cite{yu2017automatic} & Proposed \\
\hline
Nb. of parameters& 429,972& 17,340,480   & 1,783,280 &  \textbf{333,936 }\\
\hline
\end{tabular}}
\end{center}
\end{table}

\section{Conclusion and discussion} \label{conclusion}
In this paper, we utilize the unrolling method to obtain a deep image segmentation network based on the proposed generalized MS model. The original regularization term in the MS model is replaced by a data-adaptive generalized regularization that can be learned. This approach endows the proposed segmentation network naturally bears mathematical interpretability. Then, we utilize the multi-grid algorithm to solve the sub-problems and implement multi-scale feature extraction within the deep learning framework, thereby enhancing both the efficiency and accuracy of image segmentation. Furthermore, the proposed segmentation network can be extended naturally to 3D image segmentation. Compared to representative segmentation networks, the experimental results demonstrate that the proposed method performs well on both 2D and 3D images, even with a small training data size.

\bibliographystyle{plain}  



\end{document}